\documentclass[iop,apj,12pt,twocolappendix]{emulateapj}
\usepackage[normalem]{ulem}
\usepackage{acronym}
\usepackage{amsfonts}
\usepackage{amsmath}
\usepackage{apjfonts}
\usepackage{booktabs}
\usepackage[usenames]{color}
\usepackage{dcolumn}
\usepackage{enumerate}
\usepackage{epsfig}
\usepackage[plainpages=false, colorlinks=true, anchorcolor=blue, linkcolor=blue, citecolor=blue, bookmarks=false]{hyperref}
\usepackage{multirow}
\usepackage{natbib}
\usepackage{scrextend}

\newcommand{\dd}{\mathrm{d}}
\newcommand{\ii}{\mathrm{i}}
\newcommand{\km}{\mathrm{km}}

\newcommand{\cf}{\textit{c.f.}~}
\newcommand{\ie}{\textit{i.e.}~}
\newcommand{\eg}{\textit{e.g.,}~}

\newcolumntype{d}[1]{D{.}{.}{#1}}

\begin{document}

\slugcomment{Draft version \today}

\title{Neutrino-Driven Convection in Core-Collapse Supernovae: High-Resolution
Simulations}

\author{David Radice\altaffilmark{1},
  Christian D. Ott\altaffilmark{1,2},
  Ernazar Abdikamalov\altaffilmark{3},
  Sean M. Couch\altaffilmark{4,5,6}, \\
  Roland Haas\altaffilmark{7},
  and
  Erik Schnetter\altaffilmark{8,9,10}}
\altaffiltext{1}{TAPIR, Walter Burke Institute for Theoretical Physics,
  Mailcode 350-17,
  California Institute of Technology, Pasadena, CA 91125, USA,
  dradice@caltech.edu}
\altaffiltext{2}{Kavli Institute for the Physics and
 Mathematics of the Universe (Kavli IPMU WPI), The University of Tokyo, Kashiwa, Japan}
\altaffiltext{3}{Department of Physics, School of Science and Technology, Nazarbayev University, Astana 010000, Kazakhstan}
\altaffiltext{4}{Department of Physics and Astronomy, Michigan State University,
East Lansing, MI 48824, USA}
\altaffiltext{5}{Department of Computational Mathematics, Science, and Engineering, Michigan State University,
East Lansing, MI 48824, USA}
\altaffiltext{6}{National Superconducting Cyclotron Laboratory, Michigan State University,
East Lansing, MI 48824, USA}
\altaffiltext{7}{Max-Planck-Institut f\"ur Gravitationsphysik, Albert-Einstein-Institut, 14476 Golm, Germany}
\altaffiltext{8}{Perimeter Institute for Theoretical Physics, Waterloo, ON, Canada}
\altaffiltext{9}{Department of Physics, University of Guelph, Guelph, ON, Canada}
\altaffiltext{10}{Center for Computation \& Technology, Louisiana State
  University, Baton Rouge, LA, USA}

\begin{abstract}
\noindent We present results from high-resolution semi-global simulations of
neutrino-driven convection in core-collapse supernovae. We employ an idealized
setup with parametrized neutrino heating/cooling and nuclear dissociation at the
shock front. We study the internal dynamics of neutrino-driven convection and
its role in re-distributing energy and momentum through the gain region. We find
that even if buoyant plumes are able to locally transfer heat up to the shock,
convection is not able to create a net positive energy flux and overcome the
downwards transport of energy from the accretion flow. Turbulent convection
does, however, provide a significant effective pressure support to the accretion
flow as it favors the accumulation of energy, mass and momentum in the gain
region.  We derive an approximate equation that is able to explain and predict
the shock evolution in terms of integrals of quantities such as the turbulent
pressure in the gain region or the effects of non-radial motion of the fluid. We
use this relation as a way to quantify the role of turbulence in the dynamics of
the accretion shock.  Finally, we investigate the effects of grid resolution,
which we change by a factor 20 between the lowest and highest resolution. Our
results show that the shallow slopes of the turbulent kinetic energy spectra
reported in previous studies are a numerical artefact. Kolmogorov scaling is
progressively recovered as the resolution is increased.
\end{abstract}

\maketitle

\section{Introduction}
\label{sec:introduction}
The gravitational collapse of the iron core to a \ac{PNS} marks the last stage
of evolution of stars with zero-age main-sequence masses in excess of $\sim 8\
M_\odot$. A small ($\sim$ few \%) fraction of the enormous amount of
gravitational binding energy released in this process ($\sim \textrm{few }
\times 10^{53}\, \mathrm{erg}$) is somehow deposited in the outer layers (with
mass-coordinate $\gtrsim 1.5\, M_\odot$) of the star and powers some of the most
energetic explosions in nature, core-collapse supernovae (CCSNe). However, the
exact details of the mechanism responsible for re-processing the available
energy, which is mostly released as neutrinos streaming out of the \ac{PNS}, are
still uncertain \citep{janka:07, janka:12b, burrows:13a, foglizzo:15}.

\acused{CCSN}

In the standard scenario the gravitational collapse of the iron core is halted
by the repulsive component of the nuclear force at densities of a few $\times
10^{14}\ \mathrm{g}\, \mathrm{cm}^{-3}$, the inner core bounces back and
launches a strong shock wave in the supersonically infalling outer part of the
iron core.  However, this initial shock wave does not propagate all the way
out of the core. Instead, it loses energy due to neutrinos and
photo-dissociation of iron-group nuclei and succumbs to the ram pressure of the
infalling outer core material within tens of milliseconds. It turns into a
stalled accretion shock at a radius of $\sim 100-200\ \mathrm{km}$. To launch an
explosion a mechanism must be operating that revives the stalled shock.

The most commonly proposed mechanism to achieve shock revival is the delayed
neutrino mechanism \citep{bethewilson:85}. In this mechanism, neutrinos are
absorbed in the ``gain'' layer behind the shock. This is thought to provide the
necessary energy to revive and accelerate the shock in a run-away process
\citep{bethe:90, burrows:93, pejcha:12a}. Whether this mechanism is the one
powering \ac{CCSN}e is still uncertain. It is now well established that for most
progenitors the mechanism does not work in spherical symmetry
\citep{ramppjanka:00, liebendoerfer:01b, thompson:03, liebendoerfer:05,
sumiyoshi:05}. However, successful explosions have been obtained in multiple
dimensions thanks to the development of non-spherical fluid instabilities such
as the \ac{SASI} \citep{blondin:03, foglizzo:07} and neutrino-driven convection
\citep{herant:95, bhf:95, janka:96, foglizzo:06}. These instabilities reduce the
critical neutrino luminosity needed for explosion in various ways (more on this
below).  Neutrino-driven convection, in particular, seems to be the instability
most commonly found for exploding or close-to-exploding models in 3D
\citep{dolence:13, murphy:13, ott:13a, couch:13b, couch:14a, takiwaki:14a,
abdikamalov:15, melson:15a, lentz:15, melson:15b}, however 3D SASI-dominated
explosions have also been reported at least in simulations employing simplified
physics \citep{hanke:13, fernandez:15a, cardall:15}.

In this context, turbulence generated by buoyancy, SASI \citep{blondin:07,
endeve:12} and/or perturbations in the accretion flow \citep{couch:13d,
mueller:15, couch:15b}, is expected to have an important role by providing
additional effective pressure support behind the shock \citep{bhf:95, murphy:13,
couch:15a, radice:15a}. At the same time, a full understanding of
neutrino-driven convection is still missing. Previous studies were limited
either because they were in 2D, (\eg \citealt{murphy:11, fernandez:14}), or
because they did not have a sufficient resolution to fully resolve the turbulent
dynamics (\eg \citealt{hanke:12, takiwaki:12, dolence:13, couch:14a, handy:14,
couch:15a, abdikamalov:15, melson:15a, cardall:15}). The former are probably
affected by artefacts related to the symmetry assumptions due to the unphysical
inverse cascade in 2D turbulence. The latter might instead be affected by
systematic errors that are difficult to quantify without a resolution study
spanning a large range of resolutions. The studies of \citet{abdikamalov:15}
and \citet{radice:15a} suggest that convection in current \ac{CCSN} simulations
is under-resolved and dominated by the so-called bottleneck effect, a phenomenon
that arises when numerical viscosity suppresses some of the non-linear
interactions of the energy cascade and results in the accumulation of kinetic
energy at large scale \citep{yakhot:93, she:93, falkovich:94, verma:07,
frisch:08}. This could result in low-resolution simulations being artificially
more prone to explosion, as also observed in previous studies \citep{hanke:12}.

In this study, we aim at increasing the understanding of the role of turbulent
neutrino-driven convection in \ac{CCSN}e and at identifying the key effects
responsible for the global dynamics of the accretion flow in a controlled
environment and with well resolved simulations. We develop a neutrino driven
convection model that is simple enough to allow us to perform 3D simulations at
unprecedented resolution, while including all of the basic physics ingredients
of a realistic \ac{CCSN} model: an accretion shock, the converging radial
geometry, gravity, neutrino cooling, and neutrino heating.

The rest of the paper is organized as follows. First, in Section
\ref{sec:methods}, we present the details of our neutrino-driven convection
model and a description of the numerical methods we employ for our numerical
investigation. The general evolution and features of our runs are discussed in
Section \ref{sec:results.general}. There, we focus on the dynamics of large
scale quantities, such as the average shock radius and entropy profiles. In
Section \ref{sec:results.convection}, we study the dynamics of convection. In
particular, we focus on the role of convection in transporting energy and
momentum through the gain region. Section \ref{sec:results.turbulence} is
dedicated to the turbulent energy cascade and to the role of turbulence in
providing an effective additional pressure support in the postshock region. We
discuss the turbulent cascade and the kinetic energy spectrum of neutrino-driven
turbulent convection in Section \ref{sec:results.cascade}. Finally, we summarize
and conclude in Section \ref{sec:conclusions}. The appendices contain additional
technical details of our model. Appendix \ref{sec:nuclear.dissociation}
describes our treatment of nuclear dissociation at the shock and Appendix
\ref{sec:standing.accretion.shock} contains the details of the construction of
our initial conditions.

Throughout this paper we use a system of units such that $G = c =
M_{\mathrm{PNS}} = 1$, $M_{\mathrm{PNS}}$ being the \ac{PNS} gravitational mass.
Where CGS values are quoted, it is to be intended that they correspond to the
fiducial case with $M_{\mathrm{PNS}} = 1.3\ M_\odot$.

\section{Methods}
\label{sec:methods}
In the following, we present the details of our approach and of the employed
numerical methods. We note that the aim of our work is not to develop a
realistic explosion model. Rather we want to construct a controlled setup
containing all of the most important ingredients present in nature and in
state-of-the-art global simulations.

\subsection{Neutrino-Driven Convection Model}
Our initial conditions describe a stalled shock in the core of a massive star at
a given radius $r_s$. We study the accretion flow in a 3D spherical wedge domain
with a $90^\circ$ opening angle. The \ac{PNS} is excised and replaced
by an inner boundary condition at a fixed radius, $r_{\mathrm{PNS}}$.

The accretion flow is described by the equations of general relativistic
hydrodynamics,
\begin{align}\label{eq:hydro}
  \nabla_\mu J^\mu = 0\,, &&
  \nabla_\nu T^{\mu\nu} = L^\mu\,,
\end{align}
where
\begin{align}
  J^\mu = \rho u^\mu\,, &&
  T^{\mu\nu} = [\rho (1 + \epsilon) + p] u^\mu u^\nu + p g^{\mu\nu}\,,
\end{align}
and $L^\mu$ is a term that we include to model neutrino heating and cooling
(see below). $\rho$, $u^\mu$, $p$, $\epsilon$ and $g^{\mu\nu}$ denote the
fluid rest-mass density, four-velocity, pressure, specific internal energy and the
spacetime metric.

The \ac{EOS} that we employ is a modified gamma-law \ac{EOS}
\begin{equation}\label{eq:eos.base}
  p = (\gamma - 1) \rho \tilde{\epsilon}\,,
\end{equation}
where $\gamma=4/3$ is appropriate for a radiation-pressure dominated gas and
$\tilde{\epsilon}$ represents the amount of specific ``thermal energy'' available
after nuclear binding energy has been removed from $\epsilon$ for dissociated
nuclei. We account for nuclear dissociation energy in a parametrized way similar
to \citet{fernandez:09a, fernandez:09b}. See Appendix
\ref{sec:nuclear.dissociation} for the details of our implementation.

The specific entropy for our equation of state is defined up to a constant, so
we exploit this to choose the zero of entropy following \citet{foglizzo:06}
\begin{equation}\label{eq:entropy}
  s = \frac{1}{\gamma - 1}\log\bigg[\frac{p}{p_1}
  \Big(\frac{\rho_1}{\rho}\Big)^\gamma \bigg],
\end{equation}
where $\rho_1$ and $p_1$ are, respectively, the initial postshock density and
pressure (see Appendix \ref{sec:standing.accretion.shock}). In this way, $s$ is
exactly zero at the location of the shock in the initial data.

The gravity of the \ac{PNS} is included, while self-gravity of the accretion
flow is neglected, \ie we use the Cowling approximation, so that the spacetime
metric is constant in time and given by
\begin{equation}\label{eq:metric}
  \dd\, s^2 = - \alpha^2(r)\, \dd t^2 + A^2(r)\, \dd r^2 + r^2 \dd \Omega^2,
\end{equation}
where $\dd\, s^2$, not to be confused with the entropy, denotes the spacetime
line element, $\dd\Omega^2$ is the line element of the two-sphere and
\begin{equation}\label{eq:lapse}
  \alpha^2 = A^{-2} = 1 - \frac{2M_{\mathrm{PNS}}}{r}.
\end{equation}

Neutrino heating and cooling is modeled using the light-bulb scheme introduced
by \citet{houck:92, janka:01} and later used in many studies of \ac{CCSN}e, the
most recent being \citet{cardall:15}. The functional form of $L^\mu$ that we use
is similar to that of \citet{fernandez:09b}, with the appropriate
general-relativistic corrections:
\begin{equation}\label{eq:lightbulb}
  L^\mu = u^\mu \mathcal{L} = u^\mu C \rho \bigg[ f_{\mathrm{heat}} \big(K
  p_1\big)^{3/2} \Big(\frac{r_s}{r}\Big)^2 - p^{3/2} \bigg] e^{-\big([s +
  s_{\mathrm{ref}}]_-\big)^2}\,,
\end{equation}
where $C$ is an overall normalization constant, $p_1$ is the post-shock
pressure, $K$ measures the strength of the heating\footnote{$K p_1$ is the
equilibrium pressure at the location of the shock when neglecting advection, \ie
in the limit of instantaneous heating and cooling.}, $r_s$ is the shock radius
and we use the notation
\begin{equation}\label{eq:negative.part}
  [X]_- = \begin{cases}
     |X|, & \textrm{if } X < 0, \\
      0,   & \textrm{otherwise.}
  \end{cases}
\end{equation}
$f_{\mathrm{heat}}$ is set to one for most simulations and when computing the
initial conditions. We run some additional models with $f_{\mathrm{heat}} = 0.9,
0.95, 1.05$, and $1.1$ (see Table \ref{tab:runs} for more details). In Equation
\eqref{eq:lightbulb} we use a Gaussian cutoff of the heating/cooling term to
avoid catastrophic cooling on the surface of the \ac{PNS} and to suppress
heating ahead of the shock. The reference entropy $s_{\mathrm{ref}}$ is chosen
to ensure that heating is not switched off when the shock expands and $s$
becomes slightly negative. In our simulations we find (empirically)
\begin{equation}
  s_{\mathrm{ref}} = \frac{1}{\gamma - 1} \ln 2
\end{equation}
to perform well and avoid any artificial suppression of the heating in the gain
region. Note that our heating prescription neglects the non-linear feedback
between accretion and neutrino luminosity. As such, our scheme might not be
appropriate in regimes where the accretion rate at the base of the flow shows
significant variations. However, it should be reasonably adequate for the study
of nearly steady-state neutrino-driven convection we perform here.

\begin{table}
  \caption{Key Simulation Parameters.}
  \label{tab:runs}
  \vspace{-1em}
  \begin{center}
  \begin{tabular}{lccc}
  \toprule
    Run &
    $f_{\mathrm{heat}}$ &
    $\Delta r\ [\mathrm{m}]$ &
    $\Delta \theta = \Delta \varphi\ [\mathrm{deg}]$ \\
  \midrule
    Ref.       & $1.0\phantom{0}$ & $3839$            & $1.8\phantom{0}$    \\
    2x         & $1.0\phantom{0}$ & $1919$            & $0.9\phantom{0}$    \\
    4x         & $1.0\phantom{0}$ & $\phantom{0}960$  & $0.45$              \\
    6x         & $1.0\phantom{0}$ & $\phantom{0}640$  & $0.3\phantom{0}$    \\
    12x        & $1.0\phantom{0}$ & $\phantom{0}320$  & $0.15$              \\
  \midrule
    20x\footnote{Run for $\simeq 60\ \mathrm{ms}$ starting from 12x at
    $t \simeq 317\ \mathrm{ms}$.}
               & $1.0\phantom{0}$ & $\phantom{0}191$  & $0.09$              \\
  \midrule
    F0.9-Ref.  & $0.9\phantom{0}$ & $3839$            & $1.8\phantom{0}$    \\
    F0.95-Ref. & $0.95$           & $3839$            & $1.8\phantom{0}$    \\
    F1.05-Ref.\footnote{Run with extended domain: $r_{\max} \simeq 825\,
    \mathrm{km}$.\label{footnote:F105}}
               & $1.05$           & $3839$            & $1.8\phantom{0}$    \\
    F1.1-Ref.\footref{footnote:F105}
               & $1.1\phantom{0}$ & $3839$            & $1.8\phantom{0}$    \\
    F1.1-2x\footref{footnote:F105}
               & $1.1\phantom{0}$ & $1919$            & $0.9\phantom{0}$    \\
    F1.1-6x\footref{footnote:F105}
               & $1.1\phantom{0}$ & $\phantom{0}640$  & $0.3\phantom{0}$    \\
    F1.1-12x\footref{footnote:F105}
               & $1.1\phantom{0}$ & $\phantom{0}320$  & $0.15$              \\
  \midrule
    1D         & $1.0\phantom{0}$ & $\phantom{0}640$  & $-$ \\
    F1.1-1D    & $1.1\phantom{0}$ & $\phantom{0}640$  & $-$ \\
  \bottomrule\vspace{-1em}
  \end{tabular}
  \end{center}
\end{table}

Neutrino heating and cooling is consistently included in the generation of the
initial conditions with $f_{\mathrm{heat}}$ set to one.  Our initial model is
uniquely identified by the \ac{PNS} radius $r_{\mathrm{PNS}}$, the initial shock
position $r_s$, the accretion rate $\dot{M}$ and the heating parameter $K$. $C$
is fixed by the condition $\upsilon^r(r_{\mathrm{PNS}}) = 0$, where $\upsilon^i$
is the fluid three-velocity.  Note that, since our \ac{EOS} is scale free, the
\ac{PNS} mass, $M_{\mathrm{PNS}}$, scales out of the problem and our results can
be applied to any \ac{PNS} mass with the proper rescaling. The results that we
quote are for the fiducial case $M_{\mathrm{PNS}} = 1.3\ M_\odot$. In
particular, the parameters used in this work are $r_{\mathrm{PNS}} = 30$
($\simeq 57\ \km$), $r_s = 100$ ($\simeq 191\ \km$) and $\dot{M} = 10^{-6}$
($\simeq 0.2\ M_\odot\, \mathrm{s}^{-1}$).  $K$ is set to $9$ and,
correspondingly the equilibrium $C$ is found to be $C = 9 \times 10^9$, which,
for our models, corresponds to a luminosity in both the electron or
anti-electron neutrinos of\footnote{The luminosity can be obtained by recasting
the heating term in Equation \eqref{eq:lightbulb} into Eq.~(28) of
\citet{janka:01}.}
\begin{equation}
  L_\nu \simeq 1.22 \times 10^{52} \left( \frac{12\ \mathrm{MeV}}{T_\nu}
  \right)^2 \frac{\mathrm{erg}}{\mathrm{s}}\,,
\end{equation}
where $T_\nu$ is the temperature at the neutrinosphere in $\mathrm{MeV}$.
$T_\nu$ needs not to be specified by our heating/cooling prescription, because
our heating prescription depends only on the total neutrino luminosity and not
separately on the neutrino number fluxes and average energies as would have been
the case for a real transport scheme. Finally, a small random perturbation with
relative amplitude $10^{-6}$ is added to the density field to break the
symmetry. The details of the construction of the initial conditions are given in
Appendix \ref{sec:standing.accretion.shock}.

An important parameter for quantifying the convective (in)stability of the
initial conditions is the Brunt-V\"ais\"ala frequency, $\Omega_{\mathrm{BV}}$, which
we write in terms of the quantity \citep{foglizzo:06}
\begin{equation}\label{eq:bvfrequency2}
  C_{\mathrm{BV}} = \frac{\gamma-1}{\gamma} g \partial_r s\,,
\end{equation}
where $g$ is the gravitational acceleration, which we approximate as
$M_{\mathrm{PNS}} / r^2$. We define
\begin{equation}\label{eq:bvfrequency}
  \Omega_{\mathrm{BV}} = \sqrt{|C_{\mathrm{BV}}|}
  \mathrm{sign}(C_{\mathrm{BV}})\,.
\end{equation}
With our convention, negative values of $\Omega_{\mathrm{BV}}$ correspond to
unstable stratification and $|\Omega_{\mathrm{BV}}|$ gives the growth rate of
radial perturbations.

In the case of \ac{CCSN}e, an additional condition for convective instability is
that the growth rate of perturbations should be high enough so that they can
reach non-linear amplitudes and become buoyant before being advected out of the
gain region by the radial background flow \citep{foglizzo:06}. This can be
quantified by measuring the ratio between the two timescales,
\begin{equation}\label{eq:foglizzo.chi}
  \chi = \int \frac{[\Omega_{\mathrm{BV}}]_-}{|\upsilon^r|}\ \dd r\,,
\end{equation}
where the integral is extended over the gain region and we have once again used
the notation of Equation \eqref{eq:negative.part}. \cite{foglizzo:06} showed
that if $\chi \gtrsim 3$ perturbations have enough time to develop large-scale
convection. Our simulations have an initial value of $\chi = 5.33$, so we
expect them to develop large-scale convection.

\subsection{Simulation Setup}
The Equations \eqref{eq:hydro} are solved on a uniform spherical grid in
flux-conservative form \citep{banyuls:97}, using the 5th order MP5 finite
difference high-resolution shock-capturing \citep{suresh:97} scheme as
implemented in the \texttt{WhiskyTHC} code \citep{radice:12, radice:14b}.
\texttt{WhiskyTHC} employs a linearized flux-split method with carbuncle and
entropy fix that makes full use of the characteristic structure of the general
relativistic hydrodynamics equations with very small numerical dissipation.

Our computational domain covers the region $57\ \mathrm{km} \lesssim r \lesssim
442\ \mathrm{km}$ ($825\ \mathrm{km}$ for models with $f_{\mathrm{heat}} \geq
1.05$), $\pi/4 < \theta < 3 \pi/4$ and $-\pi/4 < \varphi < \pi/4$.  We use
reflecting boundary conditions at the inner boundary, inflow conditions at the
outer boundary, and impose periodicity in the angular directions. To ensure a
constant accretion rate through the shock, we add artificial dissipation, using
a standard 2nd order prescription, close to the outer boundary, always outside
of the shock front, to some of the low-resolution runs. Dissipation is found not
to be necessary for the 4x, 12x, and 20x runs. In the other simulations,
instead, we find dissipation to be necessary to prevent oscillations in the
fluid quantities close to the outer boundary that, if not suppressed, can alter
the accretion rate by a few percent. The grid spacing for the reference
resolution is $\Delta r \simeq 3.8\ \mathrm{km}$, $\Delta \theta = \Delta
\varphi = 1.8^\circ$. The reference resolution is similar to the one employed in
recent radiation-hydrodynamics simulations \citep{melson:15a, lentz:15}. For the
other resolutions we refined the grid relative to the reference run by factors
$2, 4, 6, 12, 20$, \ie up to $\Delta r \simeq 190\ \mathrm{m}$ and $\Delta
\theta = \Delta \varphi = 0.09^\circ$ for the 20x run.  All of the simulations
are carried out until $\simeq 640\ \mathrm{ms}$, apart from the 20x and the
F1.1~runs. We start the 20x at $\simeq 317\ \mathrm{ms}$ from a snapshot of the
12x run and follow it for only $\simeq 60\ \mathrm{ms}$ due to its high
computational cost. We stop the F1.1-Ref., F1.1-6x, and F1.1-12x runs
at times $t \simeq 384\ \mathrm{ms}$, $t \simeq 560\ \mathrm{ms}$ and $t \simeq
566\ \mathrm{ms}$, when the shock reaches the outer boundary of the
computational domain. Finally, we perform two additional runs in spherical
symmetry at the same (radial) resolution as the 6x resolution.  The main
characteristics of our runs are summarized in Table \ref{tab:runs}.

\section{Overall dynamics}
\label{sec:results.general}

\subsection{Shock evolution}
\label{sec:results.shock}
\begin{figure}
  \includegraphics[width=0.98\columnwidth]{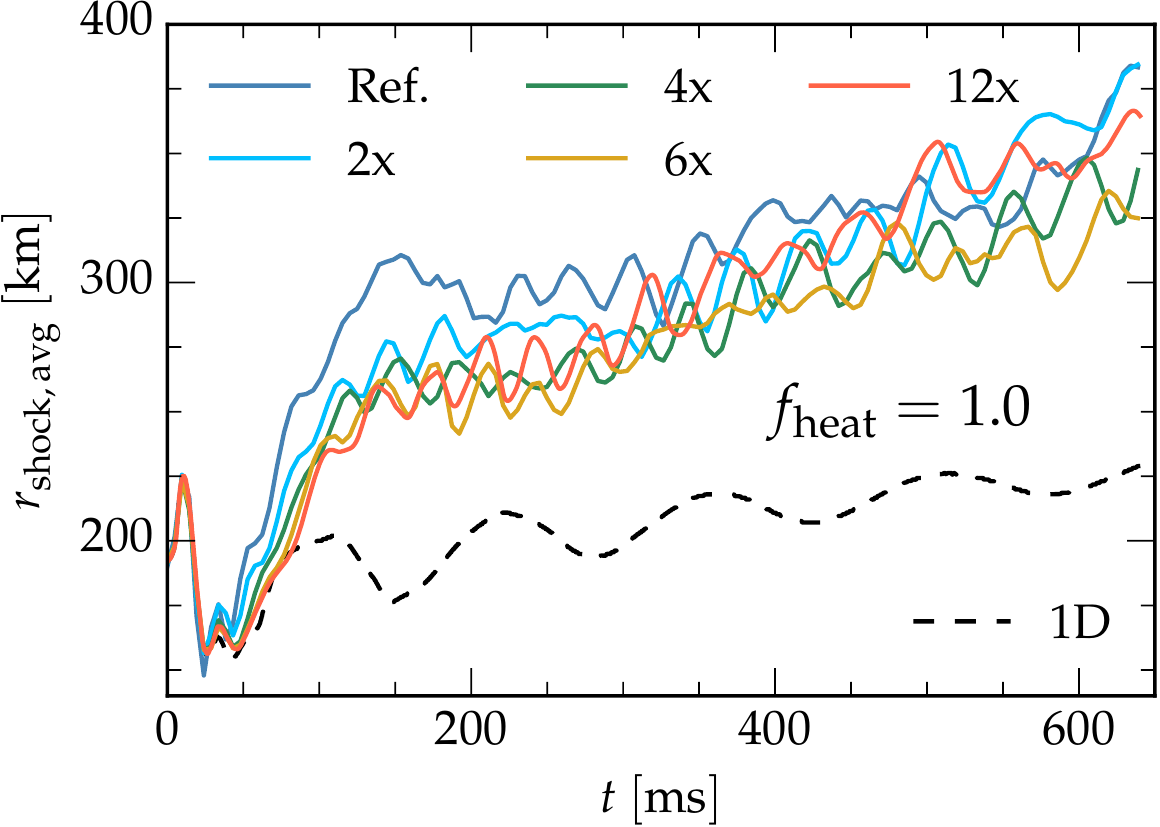}
  \caption{\label{fig:shock.radius} Average shock radius evolution for all runs
  with $f_{\mathrm{heat}}=1$.  After an initial transient, the shock radius
  expands as convection develops.  For the fiducial model, the growth slows down
  significantly and quasi-periodic oscillations appear when the convective
  plumes start to interact non-linearly with the shock front. The black dashed
  line shows the shock radius for a reference 1D run.}
\end{figure}

\begin{figure}
  \includegraphics[width=\columnwidth]{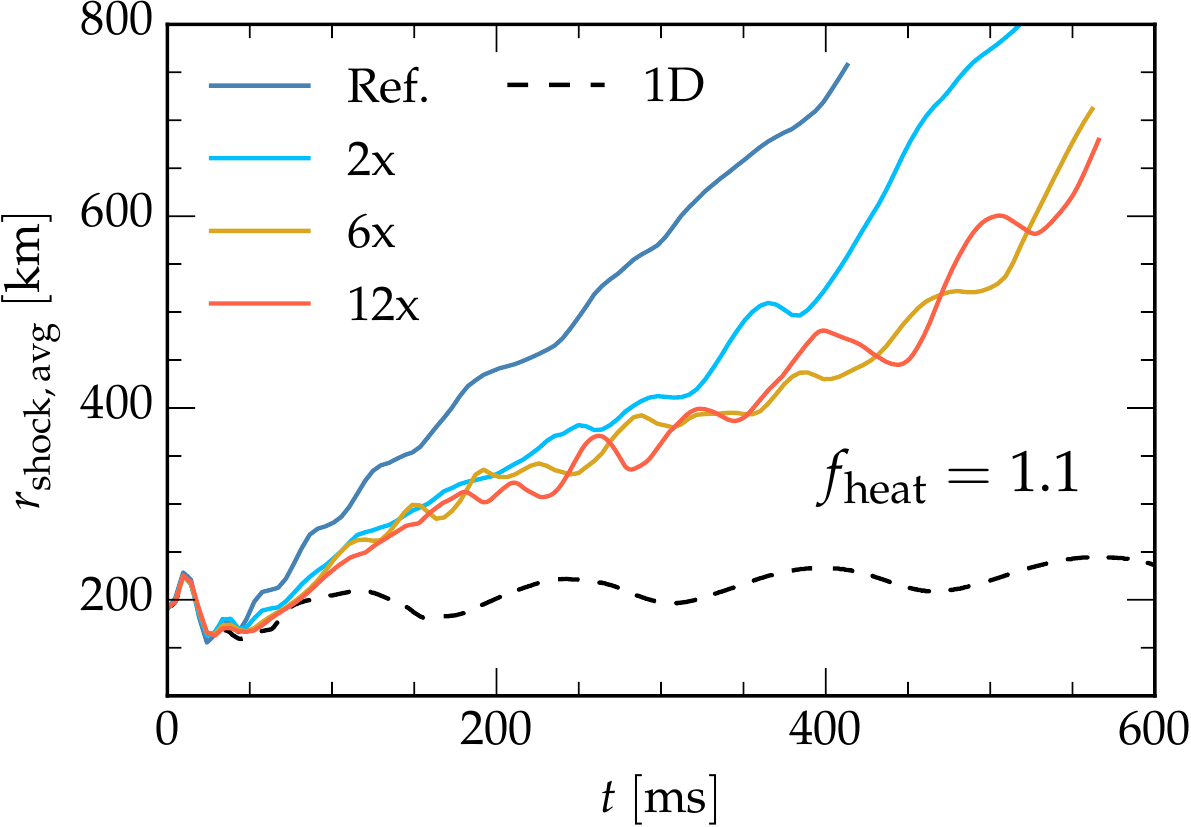}
  \caption{\label{fig:shock.radius.high.fheat} Average shock radius evolution
  for runs with enhanced heating. After the initial transient, the shock radius
  immediately starts to expand. The expansion is not significantly accelerated
  when the shock reaches large radii (as it would be in a full-physics
  simulations) partly because of our very simplified treatment of nuclear
  dissociation, with a constant specific energy loss for each fluid element
  crossing the shock. In a more realistic simulation, the amount of energy loss
  drops with radius and this leads to an accelerated expansion. The deviations
  between the reference resolution and the high resolution simulations are much
  more pronounced than for the case without enhanced heating (\cf Figure
  \ref{fig:shock.radius}). The black dashed line shows the shock radius for a
  reference 1D run.}
\end{figure}

The overall dynamics of our runs is best summarized by the average shock radius
evolution, shown in Figures \ref{fig:shock.radius} and
\ref{fig:shock.radius.high.fheat}.  The dynamics consist in an initial transient
lasting $\simeq 25\ \mathrm{ms}$ where the shock radius first expands and then
recedes. This transient is triggered by waves reflecting on the surface of the
\ac{PNS}, where the initial conditions are necessarily an approximation to the
real steady state solution, which predicts infinite density and zero velocity at
the surface of the \ac{PNS} (this is an artifact arising due to the assumption
of stationarity, see Appendix \ref{sec:standing.accretion.shock}).

For the fiducial case with $f_{\mathrm{heat}} = 1.0$, after the initial
transient, turbulence starts to develop: the initial seed perturbations trigger
the formation of small buoyant plumes at the base of the gain layer. These
plumes grow as they find their way to the shock and convection gains strength.
After $t \simeq 150\ \mathrm{ms}$ the entropy perturbations are strong enough to
cause large deformations of the shock front.

When the plumes start to interact strongly with the shock at $t \gtrsim 150\
\mathrm{ms}$, the dynamics becomes fully non-linear and characterized by a slow
growth of the shock radius and quasi-periodic oscillations with period of the
order of the advection timescale (see below). Until this point the different
runs appear to be monotonically convergent, with high-resolution simulations
having smaller average shock radii. However, as soon as the dynamics becomes
fully non-linear their shock radius evolutions lose point-wise convergence,
although the evolutionary tracks of all of the runs are broadly consistent
with each other.

By comparison, the evolution of our 1D run, also shown in Figure
\ref{fig:shock.radius}, is rather uneventful. The 1D run shows the same initial
transient as the 3D data, but afterwards it starts oscillating around its
original position and shows only a modest secular growth, which is mainly driven
by the accumulation of material in the gain region and continues for the whole
duration of the simulation. This shows that the growth of the shock radius after
$t \simeq 75\ \mathrm{ms}$ and up to $t \simeq 100\ \mathrm{ms}$ is due to the
initial development of convection, which is well captured by our runs.

The dynamics of the shock and its behavior with resolution change rather
drastically for models with enhanced heating. This can be seen in Figure
\ref{fig:shock.radius.high.fheat}, where we show the average shock radius for
the simulations with $f_{\mathrm{heat}} = 1.1$. The reference resolution
simulation starts to diverge from the 6x and 12x resolutions as soon as the
initial transient is over. The 2x resolution seems to be closer to the higher
resolution runs, which appear to be converged, but eventually also diverges away
after $t \simeq 300\ \mathrm{ms}$. Finally, the 6x and 12x resolutions appear to
be consistent with each other for the entire simulated time.

\begin{figure*}
  \includegraphics[width=\textwidth]{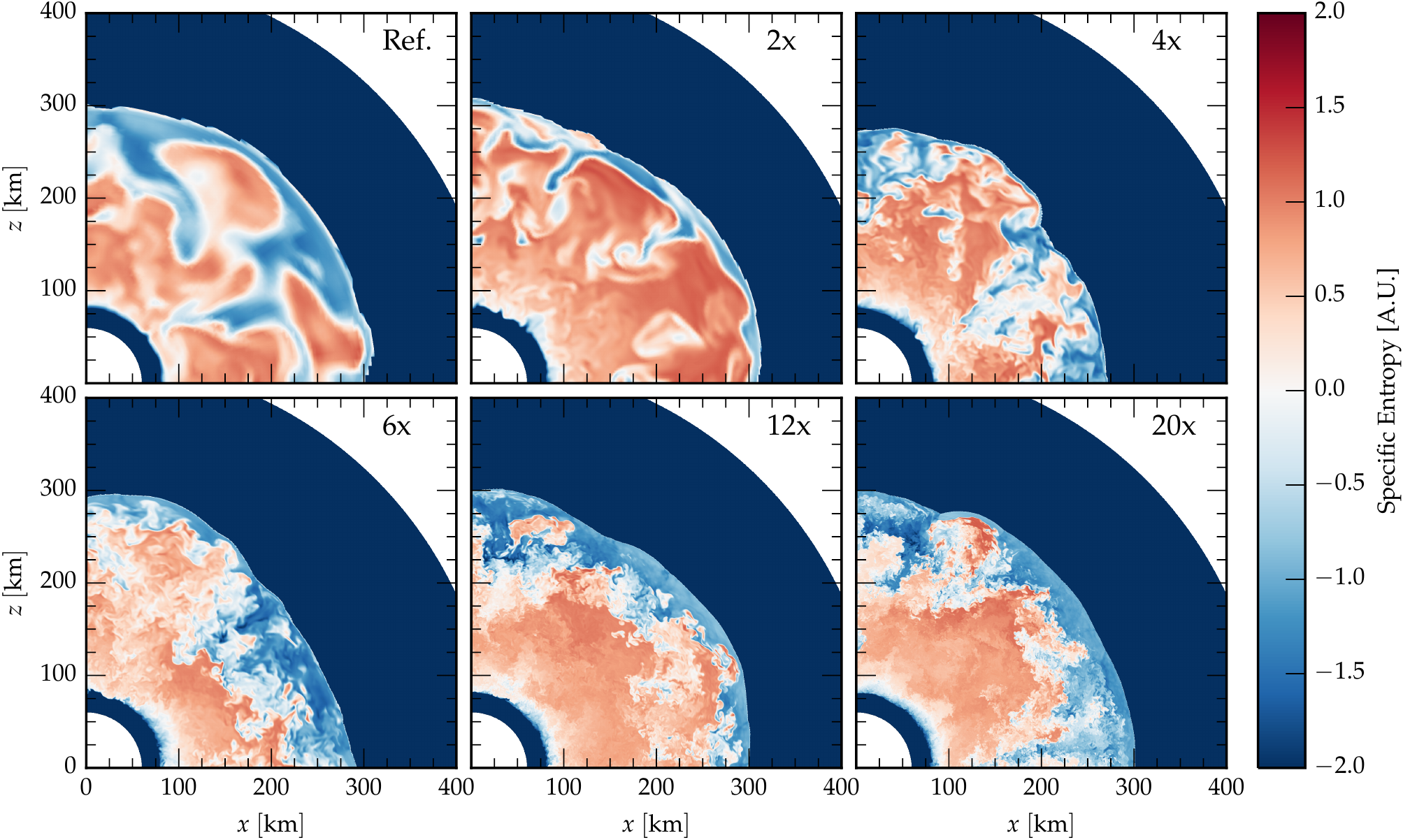}
  \caption{\label{fig:entropy.xz} Entropy, normalized as in Equation
  \eqref{eq:entropy}, in the $xz-$plane at the common time $t \simeq 365
  \mathrm{ms}$ for the simulations with $f_{\mathrm{heat}}=1.0$. Clockwise from
  the top left: reference resolution, $2x$, $4x$, $6x$, $12x$ and $20x$
  resolutions. The reference resolution is characterized by the presence of a
  few very large high-entropy plumes. As the resolution increases the topology
  of the flow becomes more complex and dominated by smaller scale features. At
  very high-resolutions plumes tend to lose coherency as small-scale turbulent
  mixing becomes effective at removing sharp features in the entropy.}
\end{figure*}

Going back to the fiducial case without enhanced heating, the qualitative
differences between resolutions are particularly evident in the visualizations
of the fluid entropy. This is shown in Figure \ref{fig:entropy.xz}, where we
display the color coded entropy in the $xz-$plane at a representative time ($t
\simeq 365\ \mathrm{ms}$). Compared with the other resolutions, the reference
resolution shows larger plumes and higher entropies. At this resolution, the
dynamics is characterized by the motion of few large structures, while, at
higher resolutions, the dynamics appears to be characterized by smaller
structures evolving on shorter timescales. Note that the appearance of large
scale coherent plumes is typically observed at the onset of explosion
\citep{dolence:13, fernandez:14, mueller:15, lentz:15}. This suggests that, as
has been also observed by \cite{hanke:12, abdikamalov:15} and consistently with
what we find for the simulations with enhanced heating, low resolution could
artificially ease the explosion (see also \citealt{couch:13b}).  On the other
hand, note that the 12x resolution, which is the highest for which we carry out
a long term evolution, is also the one showing the highest average shock radius
growth rate (Figure \ref{fig:shock.radius}), suggesting that turbulence has a
more complex role than simply destroying large-scale plumes.

As the resolution increases, first, secondary instabilities in the flow drive
down the size of the typical plumes and create more complex flow structures.
Second, at the highest resolutions (12x and 20x), plumes start to lose their
coherence due to the presence of small-scale turbulent mixing. Instead of being
characterized by entropy ``bubbles'' with sharp entropy gradients, as in the
reference resolution, the flow in high resolution simulations appears to be
dominated by the appearance and disappearance of large hot ``clouds'', \ie
entropy structures with a complex topology. An animation of the entropy
on the equatorial plane for the Ref., 2x, 4x, and 12x resolutions is included in
the online supplemental materials.

\begin{figure}
  \includegraphics[width=\columnwidth]{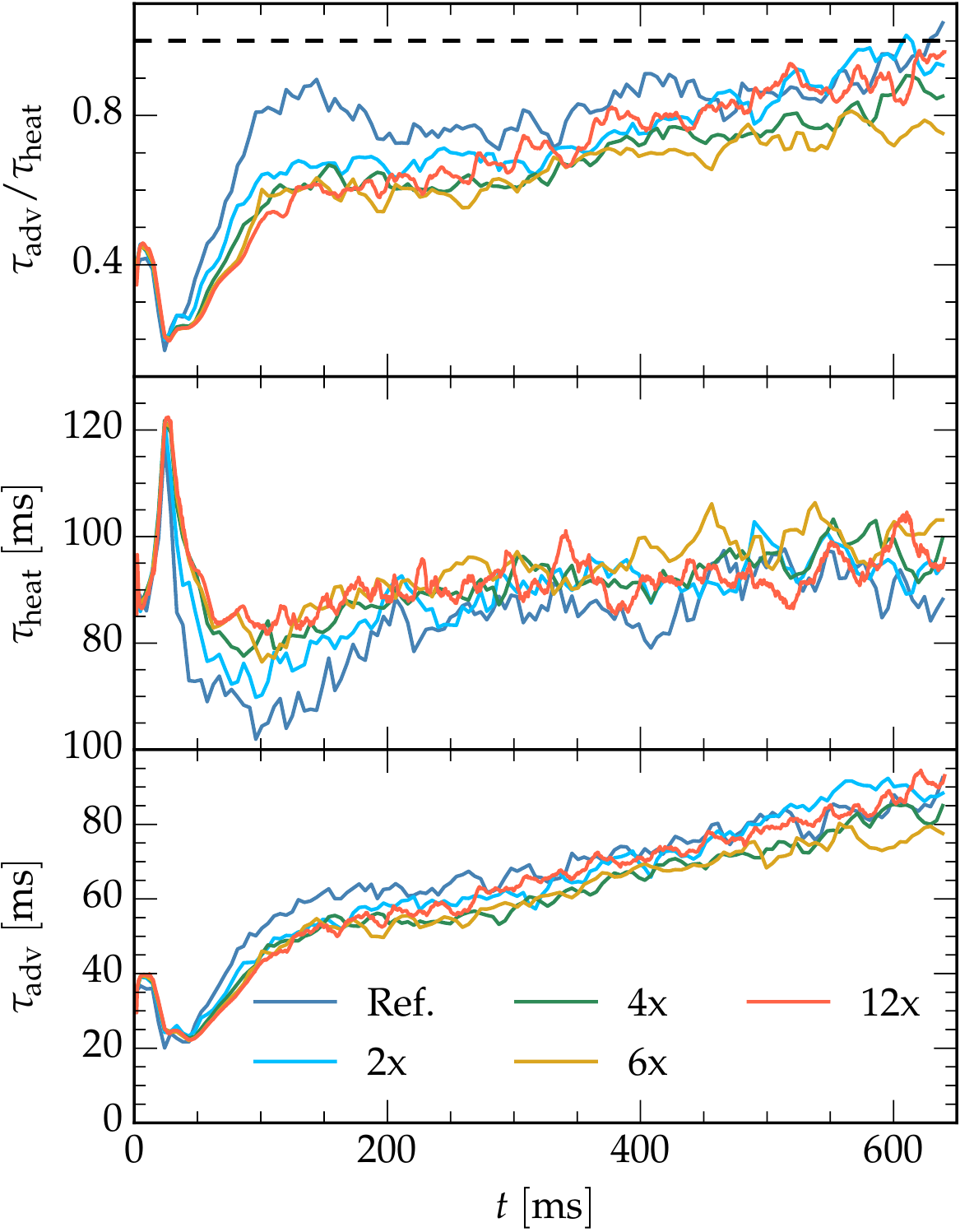}
  \caption{\label{fig:timescales} \emph{Top panel:} ratio of the advection
  timescale (Equation \ref{eq:advection.timescale}) to heating timescale
  (Equation \ref{eq:heating.timescale}). \emph{Middle panel:} heating timescale.
  \emph{Bottom panel:} advection timescale. $f_{\mathrm{heat}}=1.0$ for all of
  the runs shown in this figure. As it is the case for the shock radius
  evolution, also in the heating efficiency our simulations appear to be closely
  convergent until $t \simeq 100\ \mathrm{ms}$. Afterwards convergence is not
  monotonic, but all the runs are still in good agreement with each other,
  especially for $\tau_{\mathrm{adv}}$. The heating timescale shows a somewhat
  larger spread.}
\end{figure}

Another commonly employed diagnostic in \ac{CCSN} simulations is the ratio
between the timescale for the advection of a fluid element through the gain
layer and the time necessary for it to absorb enough energy from neutrinos to
become unbound.

Specifically, the advection timescale is typically defined as
\citep{fernandez:12, mueller:12a}
\begin{equation}\label{eq:advection.timescale}
  \tau_{\mathrm{adv}} = \frac{M_{\mathrm{gain}}}{\dot{M}}\,,
\end{equation}
where $\dot{M}$ is the accretion rate, $M_{\mathrm{gain}}$ is the total mass
in the gain region
\begin{equation}
  M_{\mathrm{gain}} = \int \rho W \sqrt{\gamma} \dd V\,,
\end{equation}
where $W$ is the Lorentz factor, $\sqrt{\gamma}$ is the spatial volume form and
the integral is extended over the gain region.

The heating timescale is defined as (\eg \citealt{fernandez:12})
\begin{equation}\label{eq:heating.timescale}
  \tau_{\mathrm{heat}} = \frac{|E_{\mathrm{bind}}|}{\dot{Q}_{\mathrm{net}}}\,,
\end{equation}
where $E_{\mathrm{bind}}$ is the binding energy of the gain region, which we
compute as in \citep{mueller:12a}:
\begin{equation}
  E_{\mathrm{bind}} = \int \big\{\alpha [\rho (1 + \tilde{\epsilon} + p/\rho)
  W^2 - p] - \rho W^2\big\} \sqrt{\gamma} \dd V\,.
\end{equation}
$\dot{Q}_{\mathrm{net}}$ is the net heating/cooling rate
\begin{equation}
  \dot{Q}_{\mathrm{net}} = \int W \mathcal{L} \sqrt{\gamma}\, \dd V\,,
\end{equation}
where $\sqrt{\gamma}$ is the determinant of the spatial metric and the integrals
in the previous equations are extended over the gain region.

The advection and heating timescales as well as their ratio for the fiducial
model ($f_{\mathrm{heat}} = 1.0$) are shown in Figure \ref{fig:timescales}. In a
similar way to what we see for the average shock radius, we find that the
different runs are monotonically convergent during the first $\sim 100\
\mathrm{ms}$. Afterwards, the various simulations have consistent trends, but
there is no point-wise convergence. After the initial transient, starting from
$t\simeq 50\ \mathrm{ms}$, the advection timescale grows by roughly a factor $2$
as convection develops. Then, starting from $t\simeq 100\ \mathrm{ms}$ the
advection timescale shows a secular growth, due to the increase of the mass in
the gain region\footnote{Note that the accretion rate is constant, so that the
advection timescale is proportional to the mass in the gain region.} and it
reaches $\sim 80\ \mathrm{ms}$ toward the end of the simulations. At the same
time, the heating timescale remains roughly constant, especially at high
resolution, and the runs slowly approach the approximate condition for
explosion, $\tau_{\mathrm{adv}} \lesssim \tau_{\mathrm{heat}}$ \citep{murphy:08,
marek:09, fernandez:12, mueller:12a}.

If we consider the ratio $\tau_{\mathrm{adv}}/\tau_{\mathrm{heat}}$ as a way to
measure the proximity of the simulations to explosion, we can see from Figure
\ref{fig:timescales} that, for the first $\sim 300\ \mathrm{ms}$, high
resolution simulations are indeed further away from explosion than low
resolution simulations as observed by \citet{hanke:12} and
\citet{abdikamalov:15} (although this trend seems to be reversed at
very high resolutions). Whether this results in explosions being triggered
artificially or not at low resolution will likely depend on how close the models
are to explosion. For instance, \citet{abdikamalov:15} found finite-resolution
effects to be small for non-exploding models and comparatively large for
exploding models. Similarly, in our simplified setup we also find the evolution
of models with enhanced heating to be more sensitive to resolution (compare
Figures \ref{fig:shock.radius} and \ref{fig:shock.radius.high.fheat}).
The recent results by \citet{melson:15b} suggest that full-physics CCSN
simulations are close to the critical threshold for explosion.  One might
speculate that, near criticality, relatively small differences as those
documented in Figure 4 could lead to dramatic consequences for some
progenitors.

\section{Dynamics of Convection}
\label{sec:results.convection}

\subsection{Convective Energy Transport}
\label{sec:results.transport}
One of the characteristics of convection is that it provides a way to transport
energy. In the context of neutrino-driven convection in \ac{CCSN}e it is
interesting to consider the role of convection in transporting energy from the
bottom of the gain layer, where neutrino deposition is the strongest, outwards,
toward the shock and, if an explosion is ultimately launched, by means of the
latter, toward the envelope of the star.

\begin{figure*}
  \includegraphics[width=\columnwidth]{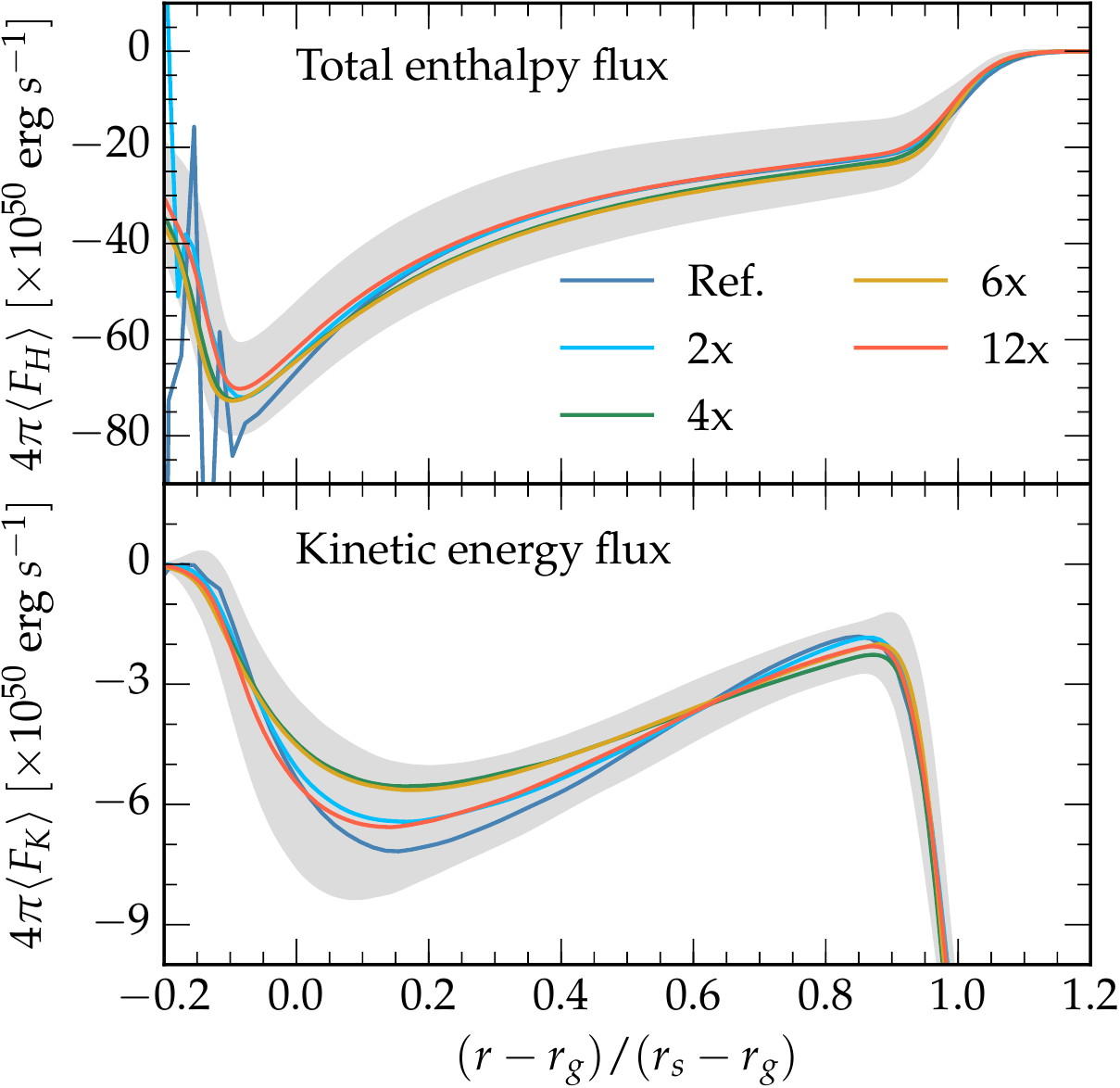}
  \hspace{0.08\columnwidth}
  \includegraphics[width=\columnwidth]{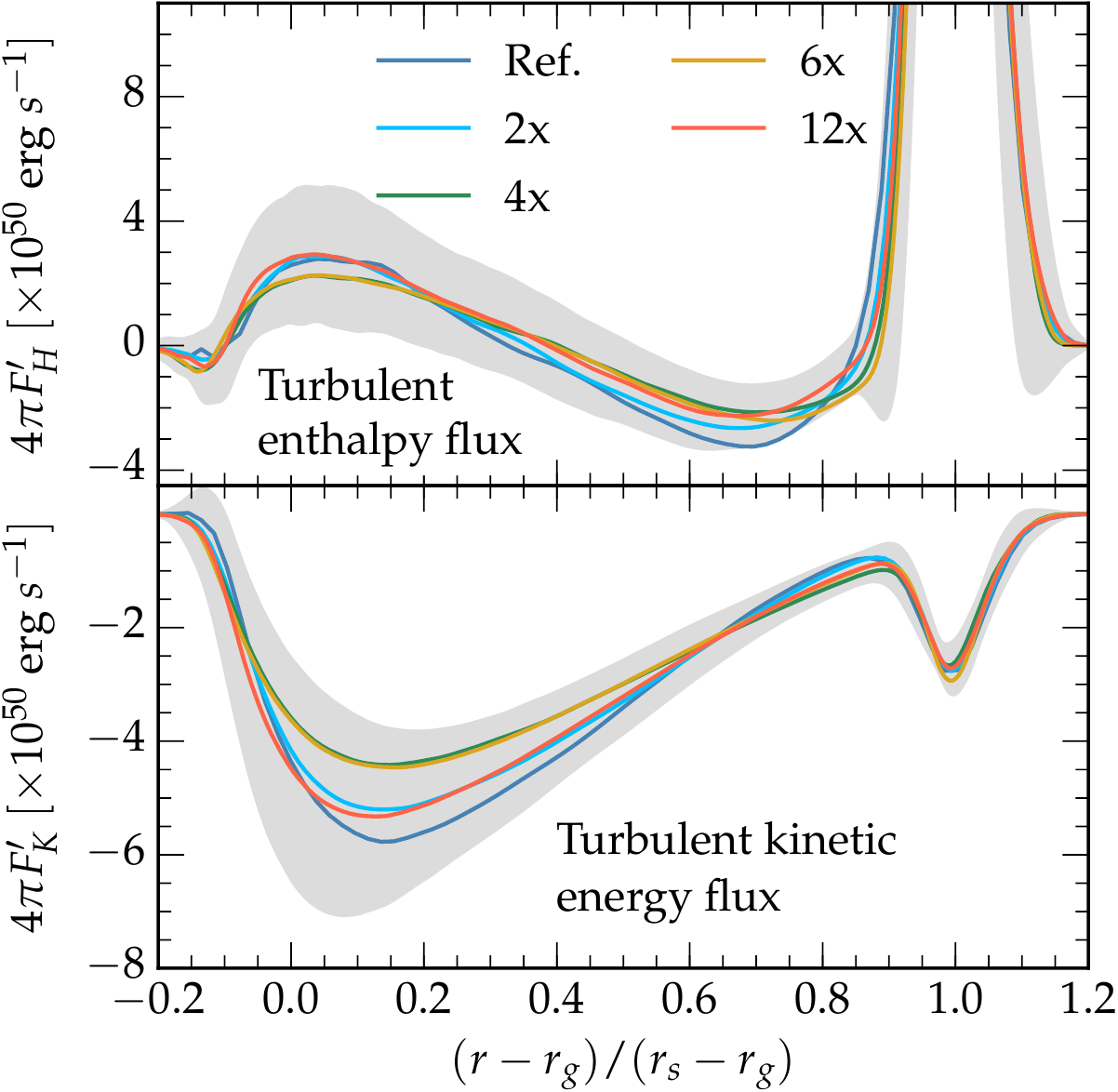}
  \caption{\label{fig:energy.fluxes} Thermal and kinetic energy transport by the
  flow for the $f_{\mathrm{heat}}=1.0$ simulations. \emph{Left panels:} time-
  and angle-averages of the kinetic and thermal energy radial energy fluxes.
  \emph{Right panels:} time- and angle-averages of the turbulent component of
  the kinetic and thermal radial energy fluxes. $r_g$ and $r_s$ are the gain and
  shock radius respectively. The time average excludes the first $t \simeq 192\
  \mathrm{ms}$ and it is carried out until the end of the simulation. The gray
  shaded region shows the standard deviation of the 12x run. Turbulence operates
  by transporting thermal energy toward larger radii and kinetic
  energy toward smaller radii.}
\end{figure*}

To analyze the efficiency of neutrino-driven convection for energy transport, we
consider the angular-integrated energy equation. Our analysis can be considered
as the general-relativistic analog of that of \cite{meakin:07b, murphy:11}, with
some minor differences. Our starting point is the angle-averaged energy equation
on the Schwarzschild background metric, Equation \eqref{eq:metric}
\begin{equation}\label{eq:energy.evolution}
  \partial_t \langle A\, r^2 E \rangle +
  \partial_r \langle r^2 [E + p] \upsilon^r \rangle =
  \mathcal{G}_E\,,
\end{equation}
where $\mathcal{G}_E$ describes heating/cooling by neutrinos and gravity
\begin{equation}
  \mathcal{G}_E = r^2 \rho W \mathcal{L} -
  \big[r^2 (E + p) \upsilon^r \big] \partial_r \log \alpha\,.
\end{equation}
The energy density is
\begin{equation}\label{eq:energy}
  E = \rho h W^2 - p\,.
\end{equation}
The total energy density can be decomposed as
\begin{equation}
  E = \rho h W (W - 1) + (\rho \epsilon + p) W + \rho W - p\,,
\end{equation}
where we can distinguish the relativistic kinetic energy density
\begin{equation}
  K = \rho h W (W - 1)\,,
\end{equation}
the ``Newtonian'' enthalpy density
\begin{equation}
  H = (\rho \epsilon + p) W\,,
\end{equation}
and the rest-mass energy density
\begin{equation}
  D = \rho W\,.
\end{equation}
The associated radial fluxes are $F_K = r^2 K\,\upsilon^r$, $F_H =
r^2 H\,\upsilon^r$ and $F_D=r^2 D\,\upsilon^r$. We can rewrite
Equation \eqref{eq:energy.evolution} as
\begin{equation}\label{eq:energy.evolution2}
  \partial_t \langle A r^2 E \rangle +
  \partial_r \langle F_K + F_H + F_D \rangle = \mathcal{G}_E\,.
\end{equation}
Furthermore, we decompose the radial velocity into a mean part and a
``turbulent'' part as
\begin{equation}
  \upsilon^r = \langle \upsilon^r \rangle + \delta\upsilon^r\,.
\end{equation}
More in general we define the turbulent velocity to be
\begin{equation}\label{eq:turb.velocity}
  \delta\upsilon^i = \upsilon^i - \langle \upsilon^r \rangle
  \delta_{r}^{\phantom{r}{i}}\,.
\end{equation}
Note that our definition of turbulent velocity is not the standard definition in
the turbulence literature, since $\langle \delta \upsilon^{\theta} \rangle$ and
$\langle \delta \upsilon^\phi \rangle$ are not necessarily exactly
zero. On the other hand, since we consider only non-rotating models, it is
natural to consider any non-radial fluid motion to be related to turbulence.
Moreover, since our background model is spherically symmetric, we expect
the ensemble averages of $\delta \upsilon^\phi$ and $\delta \upsilon^\theta$ to
vanish.

In the same way, we can split the fluxes into a mean and turbulent component as
\begin{equation}
  F_u = \bar{F}_u + F_u'\,,
\end{equation}
where $u = K, H$ or $D$, and
\begin{align}
  \bar{F}_u = \langle r^2 u \rangle \langle \upsilon^r \rangle\,, &&
  F_u' = \langle r^2 u \delta\upsilon^r \rangle\,.
\end{align}
Finally, the equation for the energy transported by convection can be written as
\begin{equation}\label{eq:energy.transport}
  \partial_t \langle A r^2 E \rangle +
  \partial_r [\bar{F}_K + \bar{F}_H + \bar{F}_D] +
  \partial_r [F_K' + F_H' + F_D'] = \mathcal{G}_E\,.
\end{equation}

Note that $F_D$ is also the flux of the angle-averaged continuity equation
\begin{equation}\label{eq:continuity}
  \partial_t \langle A r^2 D \rangle + \partial_r \langle F_D \rangle = 0\,,
\end{equation}
so that $\bar{F}_D$ and $F_D'$ can also be interpreted as mean and turbulent
contributions to the mass transport:
\begin{equation}\label{eq:mass.transport}
  \partial_t \langle A r^2 D \rangle + \partial_r \bar{F}_D +
  \partial_r F_D' = 0\,.
\end{equation}

Combining Equation \eqref{eq:continuity} with Equation
\eqref{eq:energy.evolution2}, we obtain an equation for the energy density minus
the rest-mass energy density,
\begin{equation}\label{eq:energy.net}
  \partial_t \langle A r^2 [E - D] \rangle +
  \partial_r \langle  F_K + F_H \rangle = \mathcal{G}_E\,.
\end{equation}
The quantity $E - D$ can be considered as the generalization of the sum of the
Newtonian internal and kinetic energy densities of the fluid.

To identify the important terms in the energy equation, we study the radial
profiles of the angle-averaged mean and turbulent fluxes of Equation
\eqref{eq:energy.net}. To this end, we re-map the fluxes to be a function of the
normalized radius
\begin{equation}\label{eq:mapped.radius}
  r_\star = \frac{r - r_g}{r_s - r_g}\,,
\end{equation}
where $r_g$ and $r_s$ are the average gain and shock radius respectively and we
defined the average gain radius, $r_g$, to be the radius at which neutrino
heating becomes larger than neutrino cooling in an angle-averaged sense. This
way the extent of the gain region in terms of the re-scaled radius is $0 \leq
r_\star \leq 1$ and we do not have to worry about secular changes in the shock
radius when averaging in time. Next, we average the re-mapped fluxes using data
starting at $t = 30\,000\ M_{\mathrm{PNS}} \simeq 192\ \mathrm{ms}$ to exclude
the initial transient and the development phase of convection and include only
the later quasi-steady phase.

The result of this analysis is portrayed by Figure \ref{fig:energy.fluxes},
where we show the angle-averaged total (mean + turbulent) and turbulent kinetic
and enthalpy fluxes. In each panel, the gray shaded area shows the standard
deviation of the 12x resolution. The other runs show variations of very similar
magnitude and we do not show their standard deviations to avoid overcrowding the
figure.

The large extent of the gray region in the plots is indicative of the fact that
the angle-averaged fluxes, total and turbulent for both kinetic energy and
enthalpy, show large variations in time. These are large-scale oscillations that
correlate with the quasi-periodic oscillations we see in the shock radius
(Figure~\ref{fig:shock.radius}).

The reference resolution also shows large spatial oscillations in the cooling
layer where density and pressure have a steep gradient which is not sufficiently
resolved at low resolution. These oscillations are present also for the 2x
resolution, but are confined to a much deeper layer close to the \ac{PNS} and
disappear at higher resolution.

The angle-averaged fluxes are shown in the left panels of Figure
\ref{fig:energy.fluxes}. Both the enthalpy (top) and the kinetic (bottom) energy
fluxes are negative. This means that, despite the presence of convection, the
energetics of the flow are dominated by advection and there is no net transfer
of energy upwards from the gain region to the shock. Note that this might change
if the flow transitions to an explosion, see, \eg \citep{abdikamalov:15}. In
the stalled shock phase, however, turbulence can only act in such a way as to
decrease (in absolute value) the mean fluxes and favor the accumulation of mass
and energy in the gain region, which is a necessary condition for shock
expansion \citep{janka:01}. Finally, from the amplitude of the fluxes upstream
of the shock, we can also see that, as expected, kinetic energy is the main form
with which energy is accreted through the shock, but most of it is converted
into thermal energy by the shock.

The turbulent fluxes are shown in the right panels of Figure
\ref{fig:energy.fluxes}. The angle-averaged turbulent enthalpy flux is positive
at the base of the gain region, while the angle-averaged turbulent kinetic
energy flux is negative everywhere. This is the result of buoyant plumes driving
thermal energy upwards and displacing lower entropy gas which is pushed
downwards in a process that converts thermal energy back into kinetic energy.

An important point that we can deduce from Figure \ref{fig:energy.fluxes} is
that the total amount of energy transported by turbulence is not particularly
large compared to that of the background flow. Even at the base of the gain
region, where heating is stronger and the convective enthalpy fluxes are more
intense, the turbulent angle-average enthalpy flux is at most a few $\times
10^{50}\ \mathrm{erg}\ \mathrm{s}^{-1}$, which is only of order $10\%$ of the
total enthalpy flux. Turbulence does contribute significantly to the total
kinetic energy flux, with the turbulent angle-averaged fluxes being $\sim 80\%$
of the total, but kinetic energy is dwarfed by the thermal energy in the energy
budged downstream from the shock. Obviously, these values are specific to our
accretion model and, for instance, they change by a few percent as we vary
$f_{\mathrm{heat}}$ from $0.9$ to $1.1$. However, we do not expect qualitative
differences to appear for other models during the stalled accretion shock phase.

Despite the violence of convection and the fact that buoyant plumes impinge
violently onto the shock, the total energy fluxes are still dominated by the
radial advection flow. This shows that the larger shock radius in
multi-dimensional simulations with respect to one-dimensional simulations is not
mainly due to the \emph{direct} transport of energy by convection,
which has a measurable, but small overall impact.  Turbulence, instead,
acts in a more indirect way by slowing down the drain of energy from the region
close to the shock.

This effect is analogous to, but distinct from, another well known consequence
of neutrino-driven convection: the enhancement of the absorption efficiency due
to the increased dwelling time of fluid elements in the gain region
(\eg \citealt{bhf:95, murphy:08, fernandez:09b}).

We remark that the large oscillations shown in Figure \ref{fig:energy.fluxes} in
the angle averaged turbulent fluxes at the location of the shock wave are an
artefact of our decomposition arising from the fact that the angle averaged
velocity picks up values both upstream and downstream of the shock, so that the
turbulent velocity, computed according to Equation \eqref{eq:turb.velocity}, is
artificially large.  Obviously, this is only a limitation of our analysis and
nothing ``special'' happens at the location of the shock. This can be confirmed
by looking at the total fluxes in the left panels of Figure
\ref{fig:energy.fluxes}.

Finally, concerning the behavior with resolution, we see that there is no clear
monotonic trend with resolution in the fluxes. The low resolution runs (Ref.~and
2x) tend to show more vigorous convection (as measured from the magnitude of the
turbulent fluxes) than high resolution runs (4x and 6x). However, at very high
resolution (12x) convection becomes again as strong as for the low resolution
simulations.  This is consistent with the behavior of $\tau_{\mathrm{adv}} /
\tau_{\mathrm{heat}}$ shown in Figure \ref{fig:timescales}. Note, however, that,
given the large time variations of the fluxes, these differences are not at a
sufficient level to draw strong conclusions concerning the behavior with
resolution. Also, as for the timescales, the differences in the energy fluxes
with resolution might become more pronounced for models that are closer to the
explosion threshold.

\begin{figure}
  \includegraphics[width=\columnwidth]{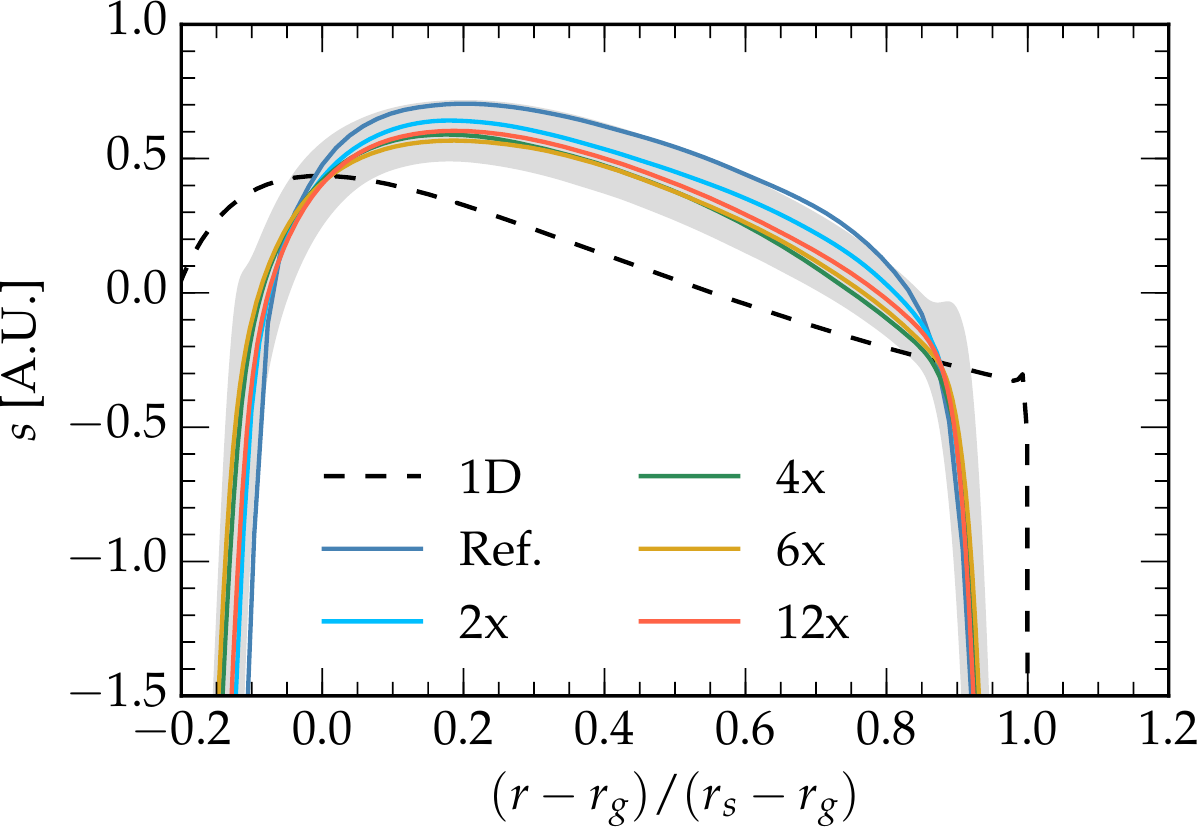}
  \caption{\label{fig:entropy} Time- and angle-averaged entropy profiles for the
  $f_{\mathrm{heat}}=1.0$ simulations. $r_g$ and $r_s$ are the gain and shock
  radius respectively. The time average excludes the first $t \simeq 192\
  \mathrm{ms}$ and it is carried out until the end of the simulation. The zero
  of the entropy is chosen according to Equation \eqref{eq:entropy}.
  Multi-dimensional convection tends to flatten the average entropy profile,
  with respect to the initial conditions (\cf Figure
  \ref{fig:init_data_entropy_omega}) or the 1D spherically symmetric simulation.
  However, convection is not efficient enough to completely cancel out the
  entropy gradient.}
\end{figure}

We note that \citet{yamasaki:06} constructed an analytic model to study the
effects of convection on the critical luminosity needed for explosion
\citep{burrows:93} and found, instead, turbulent energy transport due to
convection to have a very significant effect. The discrepancy between our
results and the model of \citet{yamasaki:06} is due to the fact that in their
model \citet{yamasaki:06} estimated the turbulent energy fluxes assuming
convection to be efficient enough to cancel the unstable gradient in the
angle-averaged radial entropy profile. This is, however, not what it is found in
simulations. The time- and angle-averaged entropy profiles from our simulations
are shown in Figure \ref{fig:entropy}. As for the fluxes, we remap the data to
be a function of $r_\star$ and then average in time. We find that
multi-dimensional convection is able to stabilize (and actually over-stabilize)
the average entropy gradient at the base of the gain region and flatten it over
most of the gain layer, as compared to the initial data (\cf Figure
\ref{fig:init_data_entropy_omega}), or to the 1D simulation. However, convection
is not efficient enough to completely remove the unstable stratification and the
average entropy profile still has a negative radial gradient over most of the
gain region. Similar entropy profiles have also been reported in other
simulations with varying degree of sophistication, (\eg \citealt{murphy:08,
hanke:12, dolence:13, ott:13a}).  This means that \citet{yamasaki:06}
overestimated the turbulent enthalpy fluxes induced by neutrino-driven
convection, which, according to simulations are not as large as to cancel the
entropy gradient. As a consequence, the model of \citet{yamasaki:06}
overestimates the effects of the turbulent energy transport on the critical
luminosity.

\subsection{Momentum transport}
\label{sec:results.turbpress}
In the light of our previous discussion, we can conclude that thermal energy
transport by turbulence appears to be only a $\sim 10\%$ effect. On the other
hand, the fact that turbulence dominates the kinetic energy balance suggests
that turbulence, and, in particular, turbulent pressure, might have a more
important role in the momentum equation. This is indeed what was already
suggested in various other studies (\citealt{murphy:13, couch:15a, radice:15a}).

\begin{figure}
  \includegraphics[width=\columnwidth]{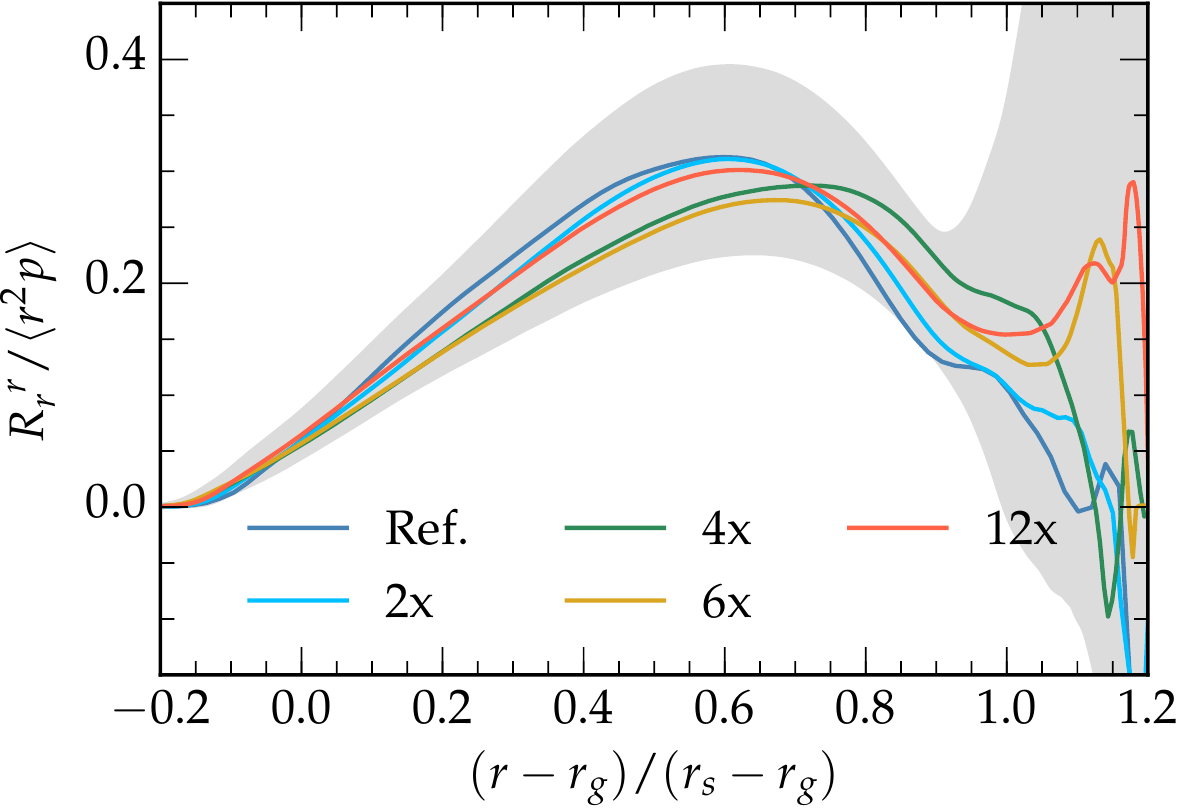}
  \caption{\label{fig:turb.press} Time- and angle-averaged ratio between the
  radial turbulent pressure and the thermal pressure for the
  $f_{\mathrm{heat}}=1.0$ runs. $r_g$ and $r_s$ are the gain and shock radius
  respectively. The time average excludes the first $t \simeq 192\ \mathrm{ms}$
  and it is carried out until the end of the simulation. Turbulence provides a
  significant contribution (of the order of $\simeq 30\%$) to the total pressure
  over most of the gain region and up to the shock, although
  $R_r^{\protect\phantom{r}r} / \langle r^2 p\rangle$ shows large variations
  close to the shock where the standard deviation of the pressure becomes of the
  order of the pressure due to the pressure jump across the shock.}
\end{figure}

To analyze this effect in detail, similarly to what we have done for the energy
Equation \eqref{eq:energy.net}, we consider the angle averaged radial momentum
equation,
\begin{equation}\label{eq:momentum}
  \partial_t \langle r^2 A S_r \rangle +
  \partial_r \langle r^2 S_r \upsilon^r \rangle = -
  \partial_r \langle r^2 p \rangle + \mathcal{G}_S\,,
\end{equation}
where the radial momentum is $S_r = \rho\, h\, W^2\, \upsilon_r$ and
$\mathcal{G}_S$ is the term containing geometric, gravitational, and neutrino
source terms in the momentum equation
\begin{equation}
\begin{split}
  \mathcal{G}_S = A^2 r^2 W^2 (\upsilon^r)^2 \mathcal{L} -
  \big[ E + p + S_r \upsilon^r \big] A^2 +
  \frac{R_{\theta}^{\phantom{\theta}\theta} +
  R_{\phi}^{\phantom{\phi}\phi}}{r} + 2 p r\,,
\end{split}
\end{equation}
where $R_i^{\phantom{i}j}$ is the Reynolds stress tensor, which we define to
be
\begin{equation}\label{eq:reynolds.stress}
  R_i^{\phantom{i}j} = \langle r^2 \rho h W^2 \upsilon_i\,
    \delta\upsilon^j \rangle.
\end{equation}
The reason for the $r^2$ factor in this definition is that this simplifies
the notation when considering spherically averaged equations.

The flux term $F_S = \langle r^2 S_r\, \upsilon^r \rangle$ in the LHS of
\eqref{eq:momentum} can also be decomposed in a mean and a turbulent part
$F_S = \bar{F}_S + F_S' = \langle r^2 S_r \rangle \langle \upsilon^r \rangle +
\langle r^2 S_r\, \delta \upsilon^r \rangle$. So that the momentum equation
can be rewritten as
\begin{equation}\label{eq:momentum.turb}
  \partial_t \langle r^2 A S_r \rangle + \partial_r \bar{F}_S +
  \partial_r F_S' = - \partial_r r^2 \langle p \rangle + \mathcal{G}_S.
\end{equation}
We note that $F_S' = R_r^{\phantom{r}{r}}$ is the radial component of the
Reynolds stress tensor.

Beside gravity, the two most important components of equation
\eqref{eq:momentum.turb} are the turbulent pressure $R_r^{\phantom{r}{r}}$ and
the thermal pressure $\partial_r \langle r^2 p \rangle$. We find the mean
angle-averaged momentum flux $\bar{F}_S$ to be contributing only $\sim 10\%$ of
the total momentum flux. The remainder is carried by turbulence $F_S' =
R_r^{\phantom{r}{r}}$ over most of the gain region.

In Figure \ref{fig:turb.press} we show the time- and angle-averaged ratio of the
turbulent pressure $R_r^{\phantom{r}{r}}$ and the thermal pressure $r^2
\langle p \rangle$ for our runs. As for the energy fluxes, we time-average the
data starting at $t \simeq 192\ \mathrm{ms}$ rescaling them as a
function of $r_\star$ (Equation \ref{eq:mapped.radius}). The shaded area shows
the standard deviation (in time) of the 12x simulation.

We find the turbulent pressure to provide roughly $\sim 30\%$ of the total
pressure support over most of the gain region and close to $\sim 20\%$ at the
location of the shock in a time-average sense. As highlighted by the shaded
region in Figure \ref{fig:turb.press}, the ratio of turbulent to thermal
pressure does, however, show significant (tens of \%) deviations in time. These
variations are particularly large close to the shock, because there the pressure
has variations of order 1 (given that the pre-shock pressure is negligible).
Turbulent pressure support drops near the base of the gain region and in the
cooling layer, where turbulence is suppressed by the strong stable
stratification near the \ac{PNS} surface.

\begin{figure}
  \includegraphics[width=\columnwidth]{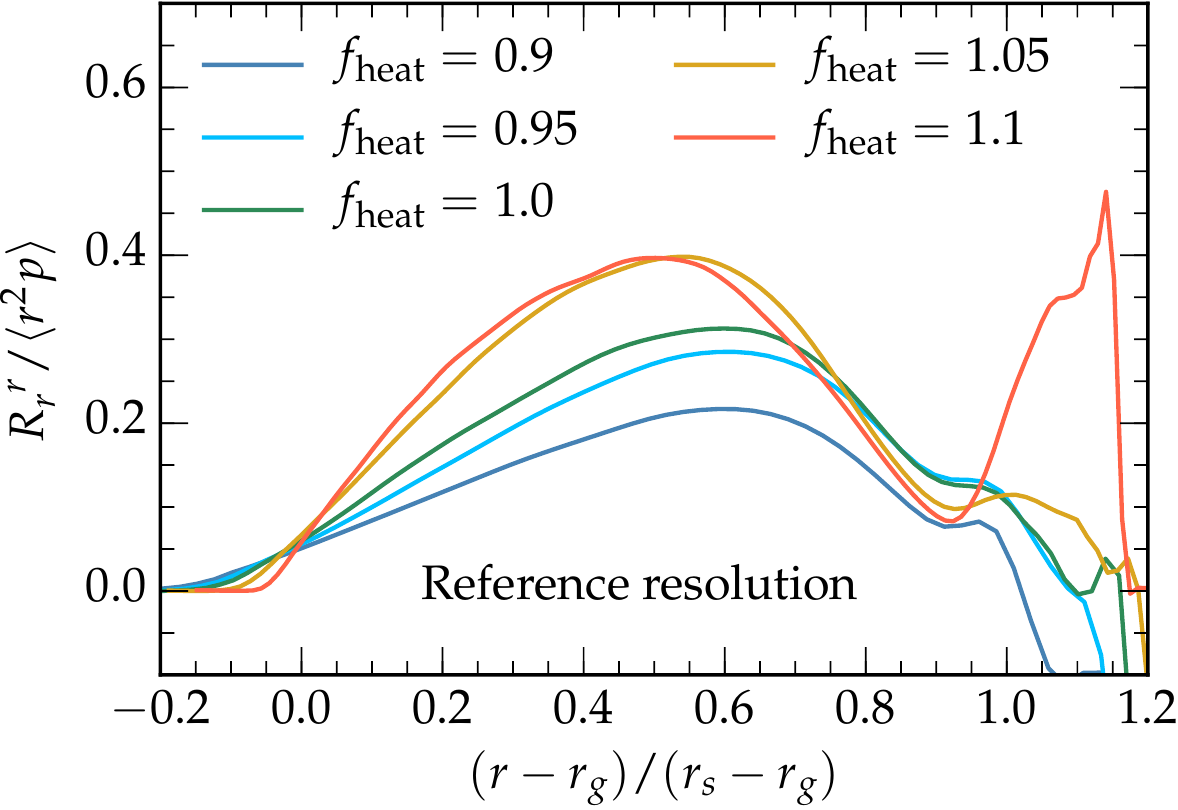}
  \caption{\label{fig:turb.press.fheat} Time- and angle-averaged ratio between
  the radial turbulent pressure and the thermal pressure as a function of the
  heating factor $f_{\mathrm{heat}}$ for the reference resolution. $r_g$ and
  $r_s$ are the gain and shock radius respectively. The time average excludes
  the first $t \simeq 192\ \mathrm{ms}$ and it is carried out until the end of
  the simulation. As the simulations approach the threshold for explosion the
  turbulent contribution to the total pressure becomes increasingly important.
  }
\end{figure}

We point out that the ratio of the effective turbulent pressure to the thermal
pressure is very sensitive to $f_{\mathrm{heat}}$: as $f_{\mathrm{heat}}$
changes from $0.9$ to $1.1$ at the reference resolution, the maximum of the
time-averaged ratio grows from $\sim 20\%$ and saturates at the $\sim 40\%$
level as shown in Figure \ref{fig:turb.press.fheat}. There we show the time- and
angle-averaged ratio of turbulent to thermal pressure for simulations with
different heating factors $f_{\mathrm{heat}}$ (see Equation \ref{eq:lightbulb})
at the reference resolution. The time-average window is the same as for Figure
\ref{fig:turb.press}. What we find is consistent with what was found by
\citet{couch:15a} who also find the ratio between turbulent pressure and
pressure to be significant. In their simulations, at the transition to
explosion, the effective pressure support from turbulence exceeds $50\%$
of the thermal pressure.

The behavior with resolution for the ratio between turbulent and thermal
pressure is in line with what we find for the energy fluxes or the
characteristic timescales. Turbulent support initially appears to decrease with
resolution, but rises again at the highest resolution (12x). Obviously, the same
caveats discussed in the case of the energy fluxes hold here. The time
variations of the ratio between turbulent pressure and pressure, with typical
amplitudes of the order of $30\%$, are such that we cannot draw strong
conclusions concerning their behavior with resolution based only on the
differences we observe. There is, however, a clear correlation between shock
radii, enthalpy fluxes and radial Reynolds stresses: large shock radii are found
in simulations with high turbulent enthalpy flux and turbulent pressure.

\begin{figure}
  \includegraphics[width=\columnwidth]{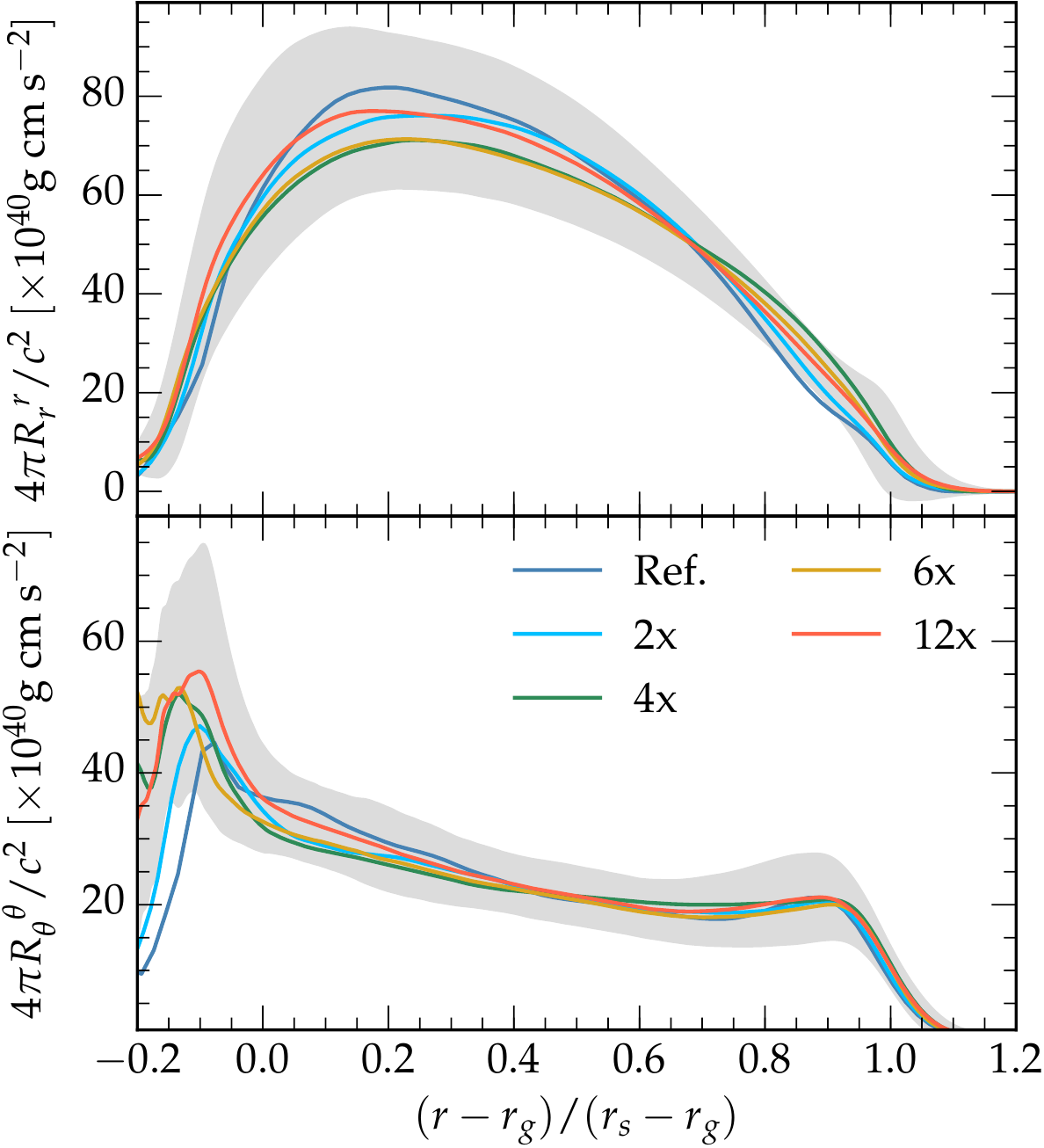}
  \caption{\label{fig:reynolds.stress} Time- and angle-averaged radial profiles
  for radial and tangential Reynolds stresses (Equation
  \ref{eq:reynolds.stress}) for the $f_{\mathrm{heat}} = 1.0$ runs. The Reynolds
  stresses show large time and spatial deviations from their average value. The
  gray shaded region shows the standard deviation of the 12x run. Turbulence is
  anisotropic with $R_r^{\protect\phantom{r}r} \simeq 2
  R_{\theta}^{\protect\phantom{\theta}\theta}$ over most of the gain region.}
\end{figure}

The time- and angle-averaged profiles for the radial and angular components of
the Reynolds stresses are shown in Figure \ref{fig:reynolds.stress}. We show
$R_r^{\phantom{r}r}$ and $R_\theta^{\phantom{\theta}\theta}$ as a function of
the normalized radius $r_\star$ (note that since our background model is
non-rotating, $R_\phi^{\phantom{\phi}\phi} \sim
R_\theta^{\phantom{\theta}\theta}$). The turbulence is highly anisotropic with
$R_r^{\phantom{r}r} \sim 2 R_\theta^{\phantom{\theta}\theta} \sim 2
R_\phi^{\phantom{\phi}\phi}$ over most of the gain region, with the important
exception of the regions close to the shock where there is near equipartition
between $R_r^{\phantom{r}r}$ and $R_\theta^{\phantom{\theta}\theta}$ as also
observed in other simulations, see \eg \citep{murphy:13, couch:15a}. The angular
components of the Reynolds stress also become dominant in the cooling layer,
where radial motions are strongly suppressed by the steep stratification.

\section{The Effects of Turbulence}
\label{sec:results.turbulence}

We have seen that the effective pressure from turbulence can contribute a
significant fraction of the total pressure support in the gain region. However,
it is not a-priori clear how to translate this into terms of the global
evolution of a \ac{CCSN}. For example, one may ask the important question of how
much the effective pressure from turbulence contributes to the evolution of the
shock radius. We address this question in the following by means of a model
explaining the shock radius evolution in terms of measurable flow quantities in
the gain region.

\subsection{Momentum Balance Equation}\label{sec:results.turbulence.mom}
We are going to derive an equation explaining the influence of turbulence on the
shock radius evolution, which will be an extension of the approach introduced by
\citet{murphy:13}. They considered the Rankine-Hugoniot conditions for a
standing accretion shock in a supernova core. In our notation
\begin{equation}\label{eq:rankine.hugoniot.classic}
  [F_S + r^2 p]_d = [F_S + r^2 p]_u\,,
\end{equation}
where $u$ and $d$ denote upstream and downstream values, respectively. They
showed how these equations could be modified to account for the turbulent
pressure. They formally decomposed the momentum flux in a turbulent and average
part in Equation \eqref{eq:rankine.hugoniot.classic} to obtain, assuming a
purely radial accretion flow upstream from the shock,
\begin{equation}\label{eq:rankine.hugoniot.turb}
  [\bar{F}_S + R_r^{\phantom{r}r} + r^2 p]_d =  [\bar{F}_S + r^2 p]_u\,.
\end{equation}
This equation is not entirely rigorous because it uses averaged quantities
inside a non-averaged equation\footnote{Note that, for instance, the averaged
equations do not formally have shocks in their solution, because the angle
average smooths all of the transitions, unless the shock is perfectly
spherical.}, however it has been shown to be well reproduced by the numerical
simulations of \citet{murphy:13} and \citet{couch:15a}. In particular,
\citet{couch:15a} found that the turbulent pressure expressed in this fashion
can be up to $50\%$ of the thermal pressure, making a very significant
contribution to the momentum balance in Equation
\eqref{eq:rankine.hugoniot.turb}.

\citet{radice:15a} pointed out that the effective adiabatic index of turbulence,
which is related to the efficiency with which turbulent energy density is
converted into thermal support, is $\gamma_{\mathrm{turb}} \simeq 2$, which is
much larger than the $\gamma = 4/3$ of a radiation pressure dominated gas. This
makes turbulent energy more ``valuable'' than thermal energy in the sense that,
per unit specific internal energy, turbulent energy contributes a greater
effective pressure than thermal energy.

More recently, \citet{murphy:15} extended Equation
\eqref{eq:rankine.hugoniot.classic} using the integral form of the momentum and
energy equations with the goal of developing a new explosion condition, but they
did not include the effects of turbulence in their analysis.  See also
\citet{gabay:15} for an alternative approach for the derivation of an explosion
condition based on the use of a virial-like relation for the moment of inertia
of the accretion layer around the \ac{PNS}.

Here, we extend the approach of \citet{murphy:13} in a way similar to
\cite{murphy:15}, but with the different goal of finding a way to quantify the
effects of turbulence on the explosion and not of constructing an explosion
diagnostic, which would be inappropriate given the limitations of our model.
Similarly to \citet{murphy:15}, our starting point is Equation
\eqref{eq:momentum}. Let us consider two spheres with radius $r_1$ and $r_0$,
with $r_1 \geq r_0$. Then Equation \eqref{eq:momentum} can be integrated between
$r_0$ and $r_1$ to yield
\begin{equation}\label{eq:momentum.integrated}
\begin{split}
  \langle F_S(r_1) + r_1^2 p(r_1) \rangle = \langle & F_S(r_0) + r_0^2 p(r_0)
  \rangle + \\ &\int_{r_0}^{r_1} \langle \mathcal{G}_S \rangle \dd r -
  \int_{r_0}^{r_1} \partial_t \langle r^2 A S_r \rangle \dd r\,,
\end{split}
\end{equation}
where $\langle F_S \rangle = \bar{F}_S + R_r^{\phantom{r}r}$. Note that for
stationary solutions, in spherical symmetry and in the limit $r_0 \to r_s^-$ and
$r_1 \to r_s^+$, $r_s$ being the shock radius, Equation
\eqref{eq:momentum.integrated} reduces to Equation
\eqref{eq:rankine.hugoniot.turb}. In the spherically symmetric, but unsteady
case and in the same limit ($r_0 \to r_s^-$ and $r_1 \to r_s^+$), Equation
\eqref{eq:momentum.integrated} yields the explosion condition derived by
\citet{murphy:15}. This is so, because, in the unsteady case, one finds
\begin{equation}\label{eq:shock.velocity}
  \lim_{\epsilon \to 0^+} \int_{r_s-\epsilon}^{r_s+\epsilon} \partial_t
  r^2 A S_r = \upsilon_s r_s^2 A(r_s)
  \big[ S_r(r_s^-) - S_r(r_s^+) \big]\,,
\end{equation}
where $\upsilon_s$ is the shock velocity.

Since our goal is to derive an equation for the shock radius directly and not
an explosion condition, we proceed differently from \citet{murphy:15}. Our
starting point is the observation that in the case in which $r_1 > r_{s,\max}$,
the LHS of Equation \eqref{eq:momentum.integrated} is well approximated by the
ram pressure of a free falling gas, \ie $\langle F_S(r_1) + r_1^2 p(r_1) \rangle
\simeq F_S(r_1) \propto r_1^{-1/2}$. This suggests that an equation for $\sim
r_1^{1/2}$ can be formally derived by integrating Equation
\eqref{eq:momentum.integrated} again with respect to $r_0$. Since we are
interested in the dynamics of neutrino-driven turbulent convection, we extend
the second integral over the whole gain region and up to $r_1$. This yields
\begin{equation}\label{eq:momentum.integrated.2}
\begin{split}
  (r_1 - r_g)& \big\langle F_S(r_1) + r_1^2 p(r_1)\big\rangle = \\
  & \int_{r_g}^{r_1} \langle F_S + r^2 p \rangle  \dd r
    + \int_{r_g}^{r_1} \dd r \int_{r}^{r_1} \langle \mathcal{G}_S \rangle  \dd r' - \\
  & \qquad\qquad \int_{r_g}^{r_1} \dd r \int_{r}^{r_1}
    \partial_t \langle (r')^2 A S_r \rangle \dd r'\,,
\end{split}
\end{equation}
where $r_g$ is the gain radius. The maximum shock radius is implicitly
determined by this equation as being the smallest value of $r_1$ for which
Equation \eqref{eq:momentum.integrated.2} holds when the expressions for the
unperturbed pre-shock accretion shock momentum flux and pressure are used as the
LHS.

If $r_1$ is chosen to be the maximum shock radius $r_{s,\max}$, Equation
\eqref{eq:momentum.integrated.2} can be used to measure the relative importance
of the different terms of the momentum equation on the shock radius. In steady
state, the RHS of Equation \eqref{eq:momentum.integrated.2} contains terms
describing the influence of
\begin{enumerate}
\item the background momentum flow
\begin{equation}\label{eq:shock.evo.mom.laminar}
  \int_{r_g}^{r_{s,\max}} \bar{F}_S \dd r\,,
\end{equation}
\item thermal pressure support
\begin{equation}\label{eq:shock.evo.press}
  \int_{r_g}^{r_{s,\max}} \langle r^2 p \rangle \dd r
    + 2 \int_{r_g}^{r_{s,max}} \dd r \int_{r}^{r_{s,\max}}
    \langle p r'\rangle    \dd r'\,,
\end{equation}
\item turbulent pressure
\begin{equation}\label{eq:shock.evo.turb}
  \int_{r_g}^{r_{s,\max}} R_r^{\phantom{r}r}  \dd r\,,
\end{equation}
\item momentum deposition by neutrinos (in the approximation of our simplified
prescription)
\begin{equation}\label{eq:shock.evo.neu}
 \int_{r_g}^{r_{s,\max}} \dd r \int_{r}^{r_{s,\max}} A^2 (r')^2 W^2
    (\upsilon^r)^2 \mathcal{L} \dd r'\,,
\end{equation}
\item gravity
\begin{equation}\label{eq:shock.evo.grav}
 - \int_{r_g}^{r_{s,\max}} \dd r \int_{r}^{r_{s,\max}}
     [E + p + S_r \upsilon^r] A^2  \dd r'\,,
\end{equation}
and
\item centrifugal support
\begin{equation}\label{eq:shock.evo.rotation}
 \int_{r_g}^{r_{s,max}} \dd r \int_{r}^{r_{s,\max}}
    \frac{R_{\theta}^{\phantom{\theta}\theta} +
    R_{\phi}^{\phantom{\phi}\phi}}{r'}\dd r'\,.
\end{equation}
\end{enumerate}
We denote the sum of all of these terms as
\begin{equation}\label{eq:shock.evo.rhs}
  \mathcal{F}(t) = \int_{r_g}^{r_1} \langle F_S + r^2 p \rangle  \dd r
    + \int_{r_g}^{r_1} \dd r \int_{r}^{r_1} \langle \mathcal{G}_S \rangle  \dd r'\,,
\end{equation}
while the term containing the time derivative of the momentum, which we compute
as the residual of the stationary equation, will be denoted as
\begin{equation}\label{eq:shock.evo.residual}
  \mathcal{R}(t) = - \int_{r_g}^{r_1} \dd r \int_{r_0}^{r_1}
    \partial_t \langle (r')^2 A S_r \rangle \dd r'\,.
\end{equation}
Both $\mathcal{F}(t)$ and $\mathcal{R}(t)$ are only functions of time.
$\mathcal{F}(t)$ encodes the flow of momentum across the gain region, while the
residual $\mathcal{R}(t)$ encodes time variations of the flow in the gain region
and the shock velocity through Equation \eqref{eq:shock.velocity}.

If we set $r_1 = r_{s,\max}$ and use the expressions for the unperturbed
upstream accretion flow,
\begin{equation}
  g(r_{s,\max}) = (r_{s,\max} - r_g) \big[ F_S +
    r_{s,\max}^2 p\big]_{\mathrm{free fall}}\,,
\end{equation}
Equation \eqref{eq:momentum.integrated.2} can be written as a formal equation
for the shock radius
\begin{equation}\label{eq:momentum.integrated.3}
  r_{s,\max}^{1/2}(t) \sim g\big(r_{s,\max}(t)\big) =
    \mathcal{F}(t) + \mathcal{R}(t)\,.
\end{equation}

\begin{figure*}
  \includegraphics[width=\columnwidth]{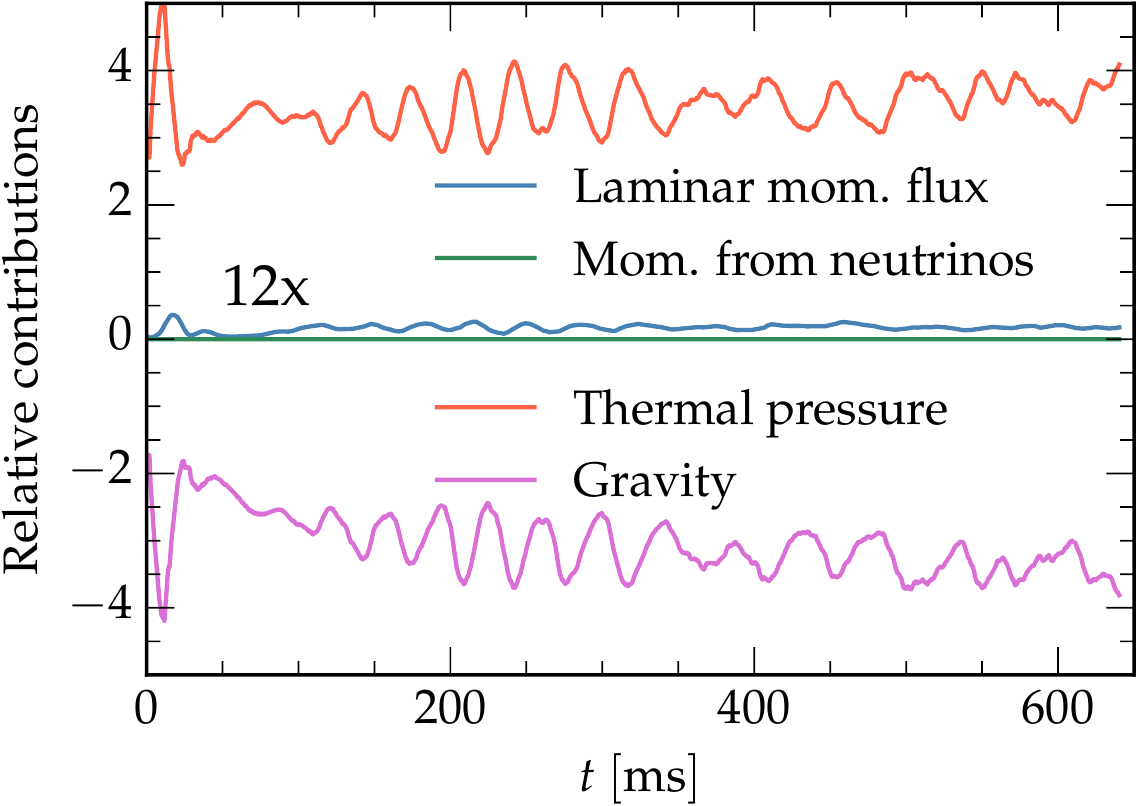}
  \hspace{0.07\columnwidth}
  \includegraphics[width=\columnwidth]{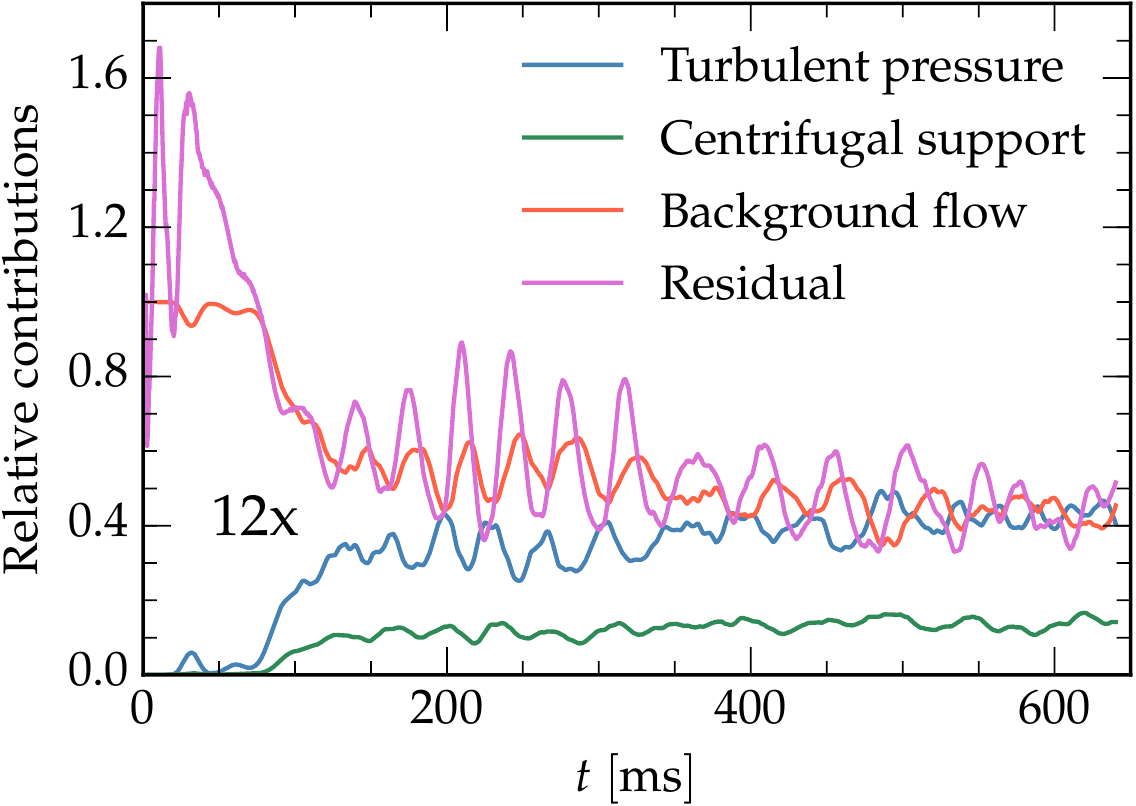}
  \caption{\label{fig:shock.radius.equation} Relative importance of the
  different terms in the equation for the radius evolution for the 12x run
  ($f_{\mathrm{heat}} = 1.0)$. \emph{Left panel:} relative contribution of the
  laminar part of the momentum flux (Equation \ref{eq:shock.evo.mom.laminar}),
  the momentum deposition by neutrinos (Equation \ref{eq:shock.evo.neu}), the
  thermal pressure (Equation \ref{eq:shock.evo.press}) and gravity (Equation
  \ref{eq:shock.evo.grav}) to the RHS of the shock radius Equation
  \eqref{eq:momentum.integrated.2}.  \emph{Right panel:} relative contribution
  of turbulent pressure (Equation \ref{eq:shock.evo.turb}), centrifugal support
  (Equation \ref{eq:shock.evo.rotation}), the background flow (the sum of all of
  the quantities in the left panel) and the residual of the stationary equation
  representing time variations of the flow (the last term of Equation
  \ref{eq:momentum.integrated.2}; Equation \ref{eq:shock.evo.residual}) to the
  shock radius Equation \eqref{eq:momentum.integrated.2}. All of the terms are
  normalized by $\mathcal{F}$ (Equation \ref{eq:shock.evo.rhs}). Gravity and
  thermal pressure give the greatest contributions, however, since the flow is
  quasi-stationary they tend to balance out. Their combined effect is of the
  same order as the turbulent support.}
\end{figure*}

We show all terms of the RHS of Equation \eqref{eq:momentum.integrated.3} for
the 12x run in Figure \ref{fig:shock.radius.equation}. All of the terms are
normalized by the sum of all terms, $\mathcal{F}$. For the purpose of making the
figure easier to read, we separate these terms into two groups: terms associated
with the ``background'' flow, shown in the left panel of the figure, and terms
associated with the turbulent motion of the fluid, shown in the right panel.

Thermal pressure and gravity are the two largest terms in magnitude in Equation
\eqref{eq:momentum.integrated.3}. The laminar part of the momentum flux
(Equation \ref{eq:shock.evo.mom.laminar}) gives only a minor contribution and
momentum deposition by neutrinos (Equation \ref{eq:shock.evo.neu}) is
unsurprisingly negligible. Note that this does not mean that neutrino heating is
negligible, but only that the direct deposition of momentum by neutrinos is
small. The fact that thermal pressure term dominates in Figure
\ref{fig:shock.radius.equation} is not in contrast with what we find for the
ratio of turbulent pressure to thermal pressure (Figure \ref{fig:turb.press}).
Namely that turbulent pressure contributes significantly to the pressure balance.
The reason for this apparent discrepancy is that most of the pressure support to
the flow in Equation \eqref{eq:shock.evo.press} comes from the bottom of the
gain region, where turbulent pressure is only $\sim 5\%$ of the thermal
pressure.

One of the first things that one notes from Figure
\ref{fig:shock.radius.equation} is that pressure and gravity nearly cancel each
other. This means that, as a very first approximation, the flow can be
considered to be quasi-stationary. The cancellation between pressure and gravity
is, however, far from being exact, as can be seen from the red graph  in the
right panel of Figure \ref{fig:shock.radius.equation}, which summarizes the net
effect of all background flow terms. This implies that, although the background
flow is quasi-stationary, it is not static, but undergoes a secular evolution
(mainly driven by the accumulation of mass and energy in the gain region;
Section \ref{sec:results.general}). The presence of secular changes in the flow
is confirmed by the fact that $\mathcal{R}(t)$, which measures the rate of
change of the total momentum of the flow, always has a positive sign and
oscillates around a constant value for most of the simulation.

Turbulent convection can be seen as a perturbation on top of this slowly
evolving background. The amplitude of the terms associated with turbulence in
the right panel of Figure \ref{fig:shock.radius.equation} is small if compared
to that of those associated with gravity and thermal pressure. This means that
the turbulent eddies are not strong enough to drastically alter the overall
settling of the accretion flow. However, turbulent fluctuations are rather large
on the scale associated with that of the secular changes of the accretion flow,
as can be seen from the fact that turbulence terms in the right panel of Figure
\ref{fig:shock.radius.equation} are as large as the residual or the net
background flow.

This is not too surprising in the light of our discussion in Section
\ref{sec:results.convection} on the energy and momentum equations. There we
showed that turbulence produces large scale changes in the energy and momentum
fluxes in space and time, but that the advection flow still dominates the
overall energetics of the flow. Figure \ref{fig:shock.radius.equation} provides
a more quantitative and well defined way to measure this contribution.

Finally, it is worth pointing out that centrifugal support from non-radial
motion produced by turbulence also provides a significant contribution to the
dynamics of the flow. It provides of order $\sim 25\%$ of the turbulent pressure
support. This is a factor that has been neglected in previous studies.

\subsection{A Model for the Shock Evolution}
So far, we have constructed a formal equation for the shock radius Equation
\eqref{eq:momentum.integrated.3} via momentum conservation. Then, we used this
equation as a way to measure the relative importance of different terms in
giving support to the shock. The following question arises naturally: is the
equation we are using merely a trivial identity involving the shock radius? Or
do the quantities $\mathcal{F}$, $\mathcal{R}$ and $r_s$ have deeper
connections?  Clearly, in the first case the analysis presented above would be
of little value. The results of our simulations suggest that this is not the
case and that $\mathcal{F}$ and $\mathcal{R}$ are relevant quantities
determining the \emph{evolution} of $r_s$.

In particular, we find that, given $\mathcal{F}$, $\mathcal{R}$, and the shock
radius at a given time, it is possible to \emph{predict} $r_{s,\mathrm{avg}}$
over a fraction of the advection timescale $\simeq 0.3 \tau_{\mathrm{adv}}$
(Equation \ref{eq:advection.timescale}) with reasonable accuracy using a simple
linear model based on Equation \eqref{eq:momentum.integrated.3}
\begin{equation}\label{eq:shock.evolution}
  \left[\frac{r_{s,\mathrm{avg}}(t + 0.3 \tau_{\mathrm{adv}})}{1\
  \mathrm{km}}\right]^{1/2} = A + B \mathcal{F}(t) + C \mathcal{R}(t)\,,
\end{equation}
where $A$, $B$ and $C$ are coefficients that we fit using a least-squares
procedure. For the 12x model, we find $A \simeq 4.84$, $B \simeq 8.54 \times
10^{-44}\, \mathrm{s}^2\ \mathrm{g}^{-1}$ and $C \simeq 1.60 \times 10^{-43}\,
\mathrm{s}^2\ \mathrm{g}^{-1}$. We find similar values also for the other
resolutions. However, we do not expect these values to be in any way universal.
Actually, we find them to change by factors of order of a few when varying the
heating factor $f_{\mathrm{heat}}$ in Equation \eqref{eq:lightbulb}.  More on
this below. Finally, we note that it is also possible to construct a similar
model for $r_{s,\max}$. We focus on $r_{s,\mathrm{avg}}$ because its time
evolution is smoother, while $r_{s,\max}$ necessarily ``jumps'' in steps that
are multiple of the grid spacing, since our analysis is not able to identify the
location of the shock to better than a single cell.

\begin{figure}
  \includegraphics[width=\columnwidth]{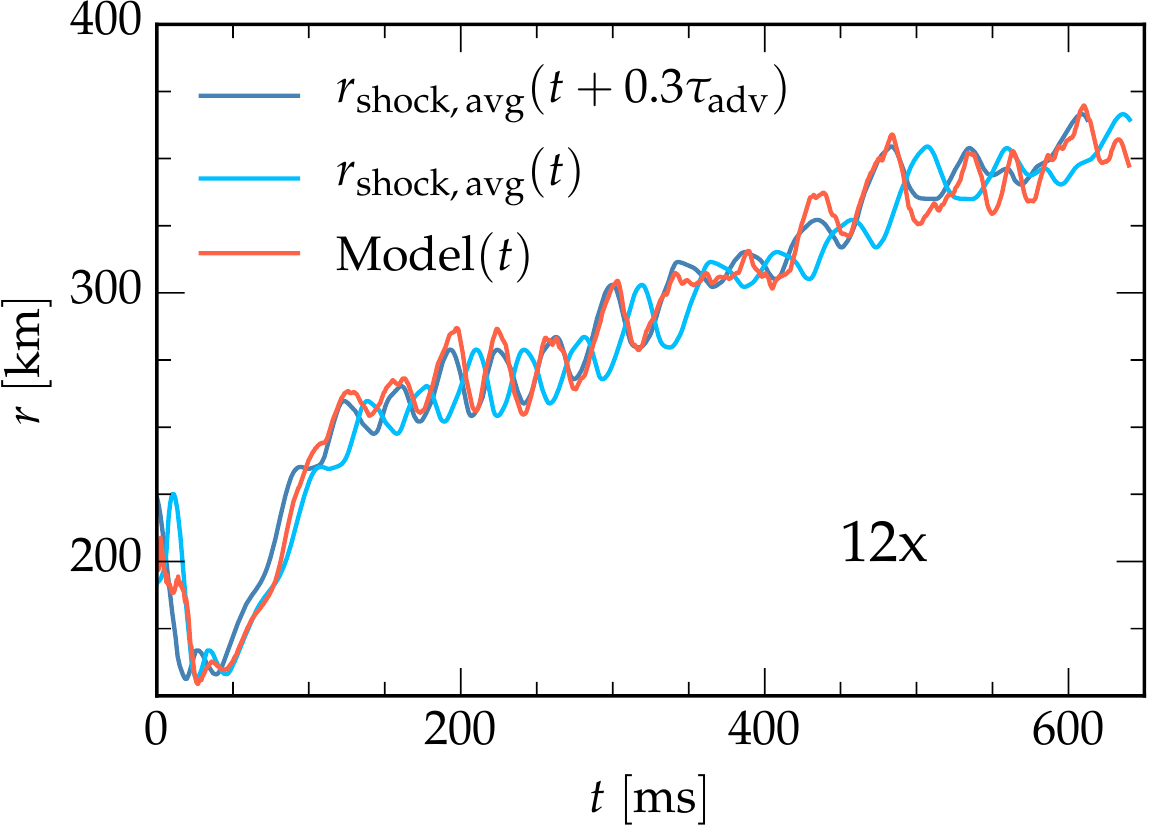}
  \caption{\label{fig.shock.radius.model} Average shock radius evolution from
  the 12x run ($f_{\mathrm{heat}}=1.0$) at the retarded time $t + 0.3
  \tau_{\mathrm{adv}}$ and its predicted value from the shock evolution Equation
  \eqref{eq:shock.evolution}.  For comparison we also show the average shock
  radius at the time when the prediction is made.  The average shock radius can
  be accurately predicted over a fraction of the advection timescale (Equation
  \ref{eq:advection.timescale}) from the knowledge of $\mathcal{F}$ (Equation
  \ref{eq:shock.evo.rhs}) and $\mathcal{R}$ (Equation
  \ref{eq:shock.evo.residual}).}
\end{figure}

The results obtained with this simple model for the shock radius are shown in
Figure \ref{fig.shock.radius.model}. There, we show the prediction for the
average shock radius at time $t + 0.3 \tau_{\mathrm{adv}}$ computed using the
data available at time $t$, $\mathrm{Model}(t)$, the actual value of the average
shock radius at the retarded time, $r_{s,\mathrm{avg}}(t + 0.3
\tau_{\mathrm{adv}})$, and the average shock radius at the time when the
prediction is made, $r_{s,\mathrm{avg}}(t)$. As can be seen from the figure, the
model specified by Equation \eqref{eq:shock.evolution} is able to predict both
the shock radius oscillations and the secular trend of the shock radius with
high accuracy. Note that we did \emph{not} include any explicit term to model
this trend in our fit: the entire shock evolution is contained in $\mathcal{F}$
and $\mathcal{R}$.

It is important to stress the fact that $\mathcal{F}$ and $\mathcal{R}$ encode
information concerning the current shock position, as well as the flow in the
gain region at the time when they are computed. They cannot be computed without
resorting to a fully non-linear simulation. In this sense, Equation
\eqref{eq:shock.evolution} is not predictive. The intriguing aspect of Equation
\eqref{eq:shock.evolution} is that it suggests that $\mathcal{F}$ and
$\mathcal{R}$ also encode information concerning the \emph{future} shock
position in a form which is easily extracted. This provides a validation to our
interpretation of the different components of $\mathcal{F}$ and of their role in
shaping the shock evolution (Section \ref{sec:results.turbulence.mom}).

It is interesting to consider whether our simple model is able to predict the
onset of a runaway explosion. This is difficult to fully assess with our
simplified simulations, because we neglect important effects leading to the
explosion, such as the sudden drop in the accretion rate following the accretion
of the $\mathrm{Si/Si\, O}$ interface \citep{buras:06b, mezzacappa:07,
ugliano:12} or asphericities in the accretion flow (\citealt{couch:13d,
mueller:15}). We also neglect the feedback of accretion on the neutrino
luminosity and we do not include all of the necessary microphysics for a fully
quantitative study.

\label{sec:shock.evo.and.fheat}
\begin{figure}
  \begin{center}
  \includegraphics[width=\columnwidth]{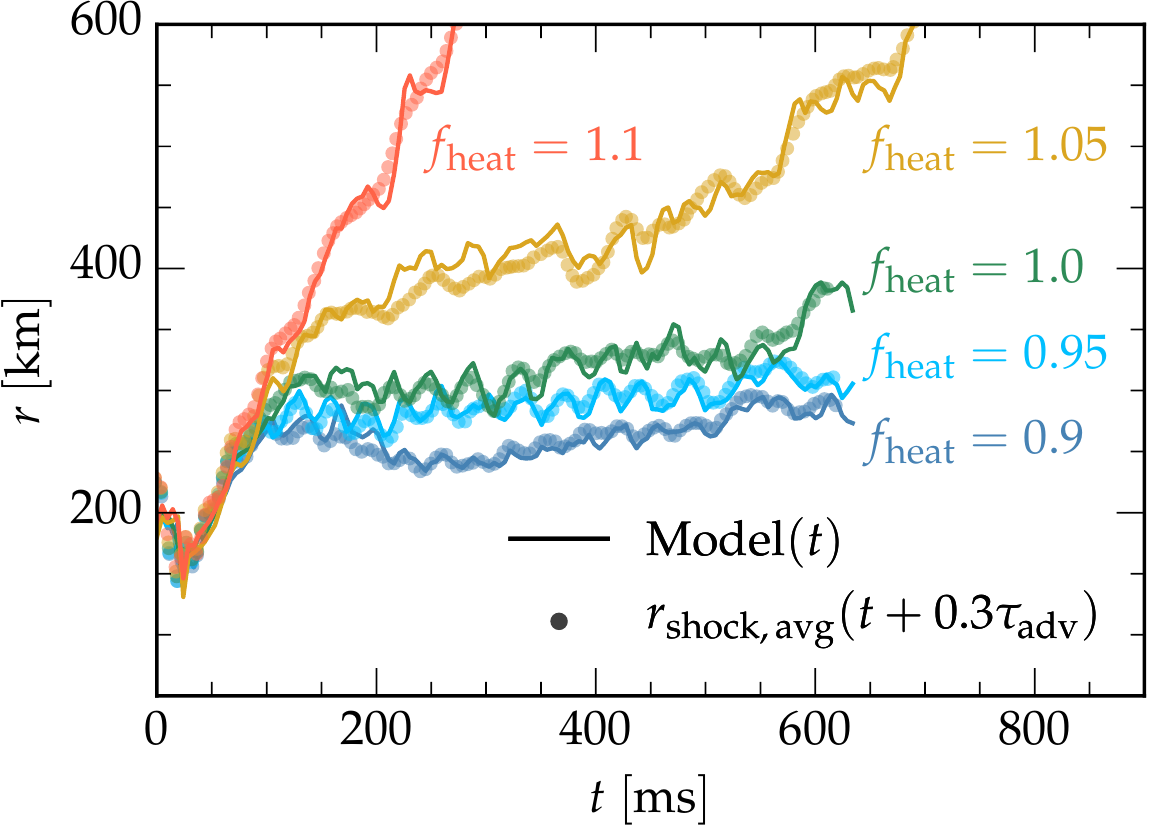}
  \end{center}
  \caption{\label{fig:shock.radius.model.fheat} Shock radius at the  retarded
  time $t + 0.3 \tau_{\mathrm{adv}}$ (dots) and its predicted value from the
  shock evolution (Equation \ref{eq:shock.evolution}) (solid lines) for the
  reference resolution, but using different heating factors.}
\end{figure}

As a first step to study the reliability of our model for exploding simulations,
we carry out a preliminary study were we obtain shock expansion / contraction by
changing the value of $f_{\mathrm{heat}}$ in Equation \eqref{eq:lightbulb} and
fit the resulting shock evolutions using Equation \eqref{eq:shock.evolution}. We
perform these simulations at the reference resolution. As anticipated above, the
fitting coefficients vary across the different runs.  Nevertheless, we find
Equation \eqref{eq:shock.evolution} to be well verified by all simulations. We
show the results of this analysis in Figure \ref{fig:shock.radius.model.fheat}.

As can be seen from this figure, the agreement between the predicted shock
radius evolution from Equation \eqref{eq:shock.evolution} and the retarded
average shock radius is reasonably good even as the heating factor is changed to
the point that the model is starting to explode. The agreement is not as good as
for the 12x model, possibly due to the higher numerical noise present in the
reference resolution data in the cooling layer and in the first few grid points
at the base of the gain region, where density and pressure increase steeply.
This noise can contaminate $\mathcal{F}$ and $\mathcal{R}$ which are the two
building blocks of Equation \eqref{eq:shock.evolution}.

\section{Turbulent Cascade}
\label{sec:results.cascade}

Following, \eg \citet{hanke:12, couch:13b, dolence:13, couch:14a, couch:15a} and
\citet{abdikamalov:15}, we consider the power spectrum of the turbulent
velocity $\delta \upsilon^i$ (Equation \ref{eq:turb.velocity}). Differently from
most previous studies, however, we do not consider the spherical
harmonics decomposition of the turbulent velocity on a sphere, but study the
actual 3D spectrum of the turbulence.

During the evolution, at each time starting from $t \simeq 192\ \mathrm{ms}$
($320\ \mathrm{ms}$ for the 20x resolution), we restrict our attention to the
largest cubic region, $B$, entirely contained in the convectively unstable gain
region and interpolate the turbulent velocity from the spherical grid used in
our simulations to a uniform Cartesian grid defined on this region. Then, we
compute the specific turbulent energy spectrum as the Fourier transform
of the two-point correlation function
\begin{equation}\label{eq:turb.energy.spec}
  \tilde{E}(k) = \frac{1}{2} \int_{\mathbb{R}^3}
  \delta(|\mathbf{k}| - k) \widehat{\delta \upsilon_i^*}(\mathbf{k})
  \widehat{\delta \upsilon^i}(\mathbf{k}) \dd \mathbf{k} \,,
\end{equation}
where $\delta(\cdot)$ is the Dirac delta, $\cdot^*$ denotes the complex
conjugation and $\widehat{\delta \upsilon^i}$ is computed as
\begin{equation}\label{eq:fourier.transform}
  \widehat{\delta \upsilon^i}(\mathbf{k}) = \int_B W(\mathbf{x}) \delta
  \upsilon^i(\mathbf{x}) \exp\left(2 \pi \ii
  \mathbf{k}\cdot\frac{\mathbf{x}}{L}\right) \dd \mathbf{x}\,.
\end{equation}
$W$ is a windowing function that smoothly goes to zero at the boundary of the
box (more on this later), $L$ is the box size, and we neglected
general-relativistic corrections in computing $\widehat{\delta \upsilon_i^*}$.
Finally, in order to account for secular oscillations in the total turbulent
energy, we normalize the spectrum to have unit integral
\begin{equation}\label{eq:turb.energy.spec.norm}
  E(k) = \left[\int_0^\infty \tilde{E}(k) \dd k\right]^{-1} \tilde{E}(k) \,,
\end{equation}
and we average in time.

Windowing in the definition of the Fourier transform
\eqref{eq:fourier.transform} is required because our data is not periodic. In
particular, $W(\mathbf{x})$ is computed as
\begin{equation}
W(\mathbf{x}) = W\left(\frac{x-x_0}{x_1-x_0}\right)
  W\left(\frac{y-y_0}{y_1-y_0}\right) W\left(\frac{z-z_0}{z_1-z_0}\right)\,,
\end{equation}
where the $(x_0, y_0, z_0)/(x_1, y_1, z_1)$ are the minimum/maximum values of,
respectively, $x,y$ and $z$ in the cube where we compute the spectra,
\begin{equation}
  W(x) = \begin{cases}
    \exp\left[\frac{-1}{1 - \left(\frac{x-\Delta}{\Delta}\right)^2}\right]
      & \textrm{if } x < \Delta\,, \\
    1 & \textrm{if } \Delta \leq x \leq 1 - \Delta\,, \\
    \exp\left[\frac{-1}{1 - \left(\frac{x-(1-\Delta)}{\Delta}\right)^2}\right]
      & \textrm{if } x > 1 - \Delta\,, \\
    0 & \textrm{otherwise}\,,
  \end{cases}
\end{equation}
and $\Delta$ is the grid spacing for the Cartesian grid, which we take to be
equal to $\Delta r$.

\begin{figure*}
  \includegraphics[width=0.98\columnwidth]{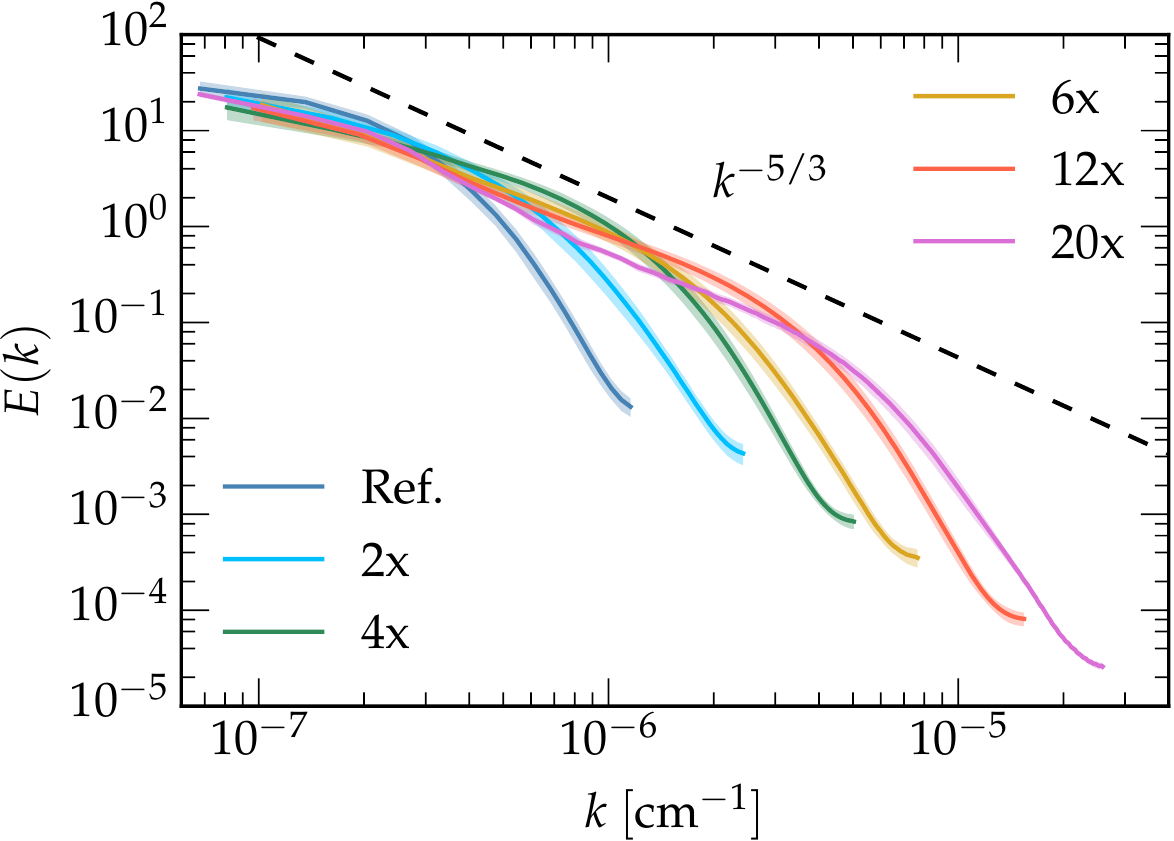}
  \hspace{0.07\columnwidth}
  \includegraphics[width=\columnwidth]{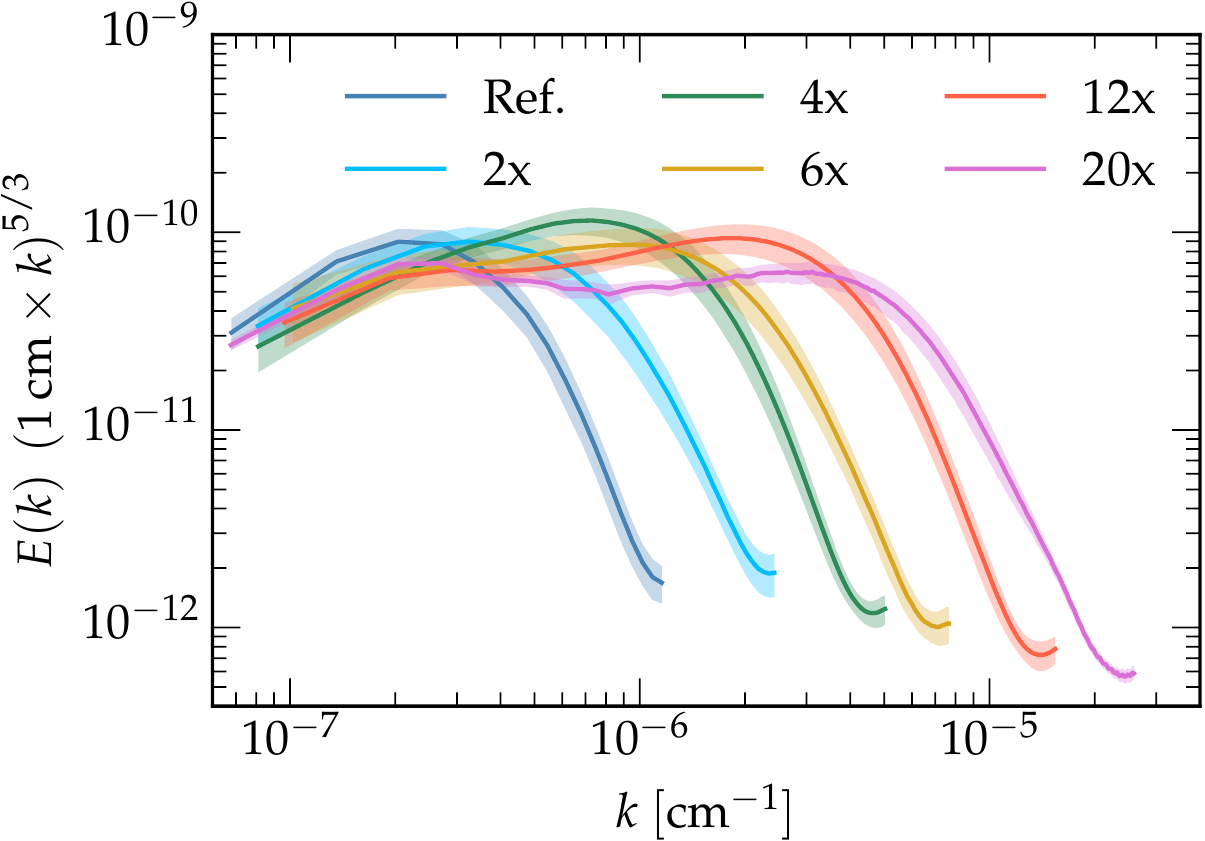}
  \caption{\label{fig:spectrum} Compensated (\ie multiplied by $k^{5/3}$)
  (\emph{right panel}) and un-compensated (\emph{left-panel}) specific turbulent
  energy power spectra, computed as in Equations \eqref{eq:turb.energy.spec} and
  \eqref{eq:turb.energy.spec.norm} for the runs with $f_{\mathrm{heat}}=1.0$.
  The width of the shaded area around each curve represents the standard
  deviation of the energy spectrum during the averaging time window. The time
  average is done from $t \simeq 192\ \mathrm{ms}$ ($320\ \mathrm{ms}$ for the
  20x resolution) until the end of the simulation.}
\end{figure*}

The normalized specific turbulent energy spectrum is shown in the left panel of
Figure \ref{fig:spectrum} for various resolutions. The right panel shows the
spectra compensated (\ie multiplied) by $k^{5/3}$ to highlight regions with
Kolmogorov scaling, which we could expect on the basis of previous
high-resolution local simulations \citep{radice:15a}. The shaded regions around
each spectrum denote the standard deviation of the spectrum during the
time-average window.

The slope of the spectra of low resolution simulations is rather shallow and
consistent with a $k^{-1}$ scaling. The low resolution spectra are comparable to
the ones reported at low-resolution by previous studies \citep{dolence:13,
couch:14a, couch:15a, abdikamalov:15}. As argued by, \citet{abdikamalov:15}
and \citet{radice:15a}, the $k^{-1}$ slope is due to the bottleneck effect
artificially trapping turbulent energy at large scale.

As the resolution increases, the spectra become progressively steeper, but even
the 12x resolution still shows a shallow $-4/3$ slope, indicative of the fact
that even at this resolution the turbulence cascade is probably dominated by the
bottleneck effect. However, the 20x resolution, which has over $1100$ points
covering the gain region in the radial direction ($\sim 15$ times more than
previous high-resolution simulations; \citealt{abdikamalov:15}), finally shows
an extended region compatible with the $k^{-5/3}$ scaling of Kolmogorov's
theory. This is particularly evident in the right panel of Figure
\ref{fig:spectrum} where we show the compensated spectrum.

This shows unambiguously that the shallow slopes reported in the \ac{CCSN}
literature are a finite-resolution effect, as previously argued by
\citet{abdikamalov:15} and \citet{radice:15a}. It also give credence to the
idea that the turbulent cascade of kinetic energy in neutrino-driven convection
is well described by Kolmogorov theory, despite the presence of non-classical
effects such as anisotropy, the geometric convergence of the flow, and mild
compressibility.

Unfortunately, the computational costs of running at the 20x resolution are
prohibitive even for our simplified setup and we could not run the 20x
simulation for more than $\simeq 60\ \mathrm{ms}$, which is roughly equal to
$\tau_{\mathrm{adv}}$. This is enough to study the energy spectrum at
intermediate and small scales (including the inertial range), which we find to
have reached a new equilibrium already $\sim 3\ \mathrm{ms}$ after the mapping
from the 12x run. It is also sufficient to serve as benchmark data for future
validation of turbulence models.  Generating such benchmark data is one of the
goals of the present work. It is not enough evolution time, however, to assess
whether the structure of the gain layer and the general dynamics of the
\ac{CCSN} simulation model changes once the inertial range starts to be
resolved.

\section{Conclusions}
\label{sec:conclusions}

Multi-dimensional instabilities are expected to play a fundamental role in the
mechanism powering most \ac{CCSN}e (\eg \citealt{murphy:11, murphy:13, hanke:13,
couch:13b, takiwaki:14a, couch:15a, melson:15a, lentz:15, melson:15b}).
Neutrino-driven convection, in particular, is most commonly associated with
post-collapse evolutions having strong neutrino heating and, in general,
conditions that are most favorable for explosion \citep{dolence:13, murphy:13,
ott:13a,  couch:13b, couch:14a, takiwaki:14a, abdikamalov:15}, however it is
not excluded that \ac{SASI} dominated \ac{CCSN}e could also explode
\citep{hanke:13, fernandez:15a, cardall:15}.

Despite its central role in the context of the delayed neutrino mechanism,
neutrino-driven convection has not been studied in a systematic way before.
Previous studies were either performed in 2D, \eg \citep{murphy:11,
fernandez:14}, or spanned a relatively small range in resolution
\citep{hanke:12, dolence:13, couch:14a, couch:15a}. Hence, it is difficult to
assess to what level they are affected by finite-resolution effects.
\citet{abdikamalov:15} and \citet{radice:15a} showed that, at the resolutions
typically used in 3D \ac{CCSN} simulations the dynamics of the turbulent cascade
of energy from large to small scale is severely affected by numerical viscosity.
This artificially prevents kinetic energy from decaying to small scales and
leads to an unphysical accumulation of energy at the largest scales, a
phenomenon known as the bottleneck effect \citep{yakhot:93, she:93,
falkovich:94, verma:07, frisch:08}. This large scale energy, in turn, results in
an additional pressure support to the accreting flow \citep{radice:15a}.
Considering the fact that turbulent pressure was found to be crucial in
triggering explosions \citep{couch:15a}, having artificially large turbulent
pressures, might result in a qualitative change in the evolution of a
simulation. For this reason, it is important to quantify finite-resolution
effects in \ac{CCSN} simulations.

In the present study, we performed a series of semi-global neutrino-driven
convection simulations with the goal of understanding the dynamics of
neutrino-driven convection and the effects of finite resolution in \ac{CCSN}
simulations. Our simulations are rather unsophisticated when compared to
state-of-the-art radiation-hydrodynamics simulations, \eg \citep{melson:15b,
lentz:15}. However, they include most of the basic physics ingredients relevant
for neutrino-driven turbulent convection and they have the advantage of being
completely under control. The converging geometry, the advection of gas through
an accretion shock toward a central \ac{PNS}, gravity, photo-dissociation of
heavy-nuclei at the shock and neutrino/heating cooling are all included in a
completely controlled way. The main limitations of our model are that we neglect
the non-linear feedback between accretion and neutrino luminosity, which we
assume to be constant, and that we fix the amount of specific energy lost to
nuclear dissociation.  These approximations would be particularly limiting in
the study of the transition to explosion of our models. Such a study would
require us to follow the shock as it develops large radial displacements with
respect to its initial conditions and correctly account for significant changes
in the accretion rate. However, this is not the aim of this work. Our current
approximations are expected to be adequate for the study of the nearly
stationary neutrino-driven convection we report here. We considered a constant
accretion rate and analytic, stationary initial conditions so as to be able to
perform long term evolutions and collect well-resolved statistics of the
turbulent flow.  We employed high-order, low dissipation numerical methods, a
grid adapted to the problem (a spherical wedge) and we varied the grid scaling
across different simulations by a factor 20, achieving unprecedented resolutions
for this kind of study, with a radial spacing of $191\ \mathrm{m}$ and an
angular resolution of $0.09^\circ$ in the gain layer.

We find that, as resolution increases, the qualitative dynamics of the flow
changes drastically. At low resolution, the dynamics is characterized by the
presence of large, slowly evolving, high-entropy plumes. At higher resolution,
the dynamics is dominated by smaller structures evolving on a faster timescale.
Given that the transition to explosion seems to be preceded by the formation of
large, long-lived, high-entropy plumes \citep{fernandez:14, mueller:15,
lentz:15}, this is a first indication that low resolution might be artificially
favouring explosion.

At high resolution, turbulent mixing is very effective at smoothing out sharp
interfaces between high and low entropy regions: high-entropy plumes lose their
coherence due to small scale mixing and they resemble more ``clouds'' than
``bubbles''.  This means that the separation of the flow into very well defined
high and low entropy regions seen in most simulations is also an artifact of low
resolution.  This is not too surprising: the physics of neutrino-driven
convection is not that of a multiphase flow. This calls into question the
usefulness of arguments describing neutrino-driven convection in terms of an
ensemble of ``bubbles'' moving through the accretion flow.

Despite these large qualitative changes with resolution, but as predicted by
\citet{radice:15a}, we find large scale quantities to be consistent among the
different resolutions for our fiducial model. In other words, we find global
quantities, such as the shock radius, the typical timescales for advection and
heating, to be consistent across all of the resolutions and to be even
monotonically convergent in the first $100\ \mathrm{ms}$, a phase in which
convection is developing, but the buoyant plumes have not yet managed to
strongly interact with the shock. Note, however, that we also find that this
picture changes drastically for models where we induce an expansion of the shock
by artificially increasing the heating rate. For these models, we find low and
high resolution simulation to be diverging after the first $50\ \mathrm{ms}$ and
low resolution simulations showing earlier shock expansion.

We also find, in agreement with \citet{hanke:12} and \citet{abdikamalov:15},
that low resolution typically yields more favorable conditions for explosion,
especially at early times. These differences are rather modest for our fiducial
set of simulations, but are more pronounced for  simulations that are closer to
or at the transition to explosion.  Given that some of the current full-physics
models appear to be on the verge of explosion \citep{melson:15b}, our results
serve as an additional reminder that a resolution study is necessary to confirm
any result. At the same time, we think, in the light of our findings, that some
cautious optimism can be justified in the sense that many quantities of interest
in \ac{CCSN}e appear to be well converged at modest resolution, even though
others, like the velocity spectra $E(k)$, are affected by serious artifacts
until very high resolution is reached.

Furthermore, it is interesting to note that our results suggest that the flow
dynamics and the resulting CCSN evolution may change quantitatively and
qualitatively at very high resolution when turbulence begins to be resolved. The
simulation at twelve times our fiducial resolution exhibits a reversal of the
just discussed trend with resolution: its shock radius evolution has the
steepest slope. Explosion diagnostics such as the ratio of advection to heating
timescales suggest that it is approaching explosion faster than less resolved
simulations. We cannot fully understand this trend until it becomes possible to
carry out even higher-resolution long-term simulations. However, we speculate
that our finding may be a consequence of the increasing non-linear coupling of a
greater range of scales and the development of strong intermittency as the
inertial range of turbulence begins to be resolved. The effects of fully
resolved turbulence (perhaps captured by a sub-grid model) may ultimately be
beneficial for explosion.

In order to better quantify the importance of turbulent convection for \ac{CCSN}
explosions, we studied the efficiency with which neutrino-driven convection
transports energy and momentum across the gain region. We find the energy
balance in the flow to be dominated by the thermal energy and the overall
energetics to be driven by the background advective flow. Turbulence opposes the
overall negative (down-flowing) radial enthalpy fluxes, but it is only able to
contribute a small $\sim 10\%$ correction to the overall thermal energy flow.
The kinetic energy evolution, on the other hand, is dominated by turbulence,
which provides $\sim 80\%$ of the kinetic energy flux and $\sim 90\%$ of the
advective part of the momentum flux (the part of the momentum flux not
containing the pressure gradient). We also find the effective pressure support
provided by turbulence to be significant and of the order of $\sim 30\%-40\%$ of
the thermal pressure in our simulations.

According to our results neutrino-driven turbulent convection plays a more
important role in the evolution of the momentum than in the evolution of the
energy. This suggest that the large differences in, \eg shock radii, between
turbulent multi-dimensional and one-dimensional simulations can be mostly
accounted for by the effects of turbulence in the momentum equation in agreement
with \cite{couch:15a}. In this respect, we showed that it is possible to derive
an equation that can explain and even predict the shock radius evolution over a
fraction of the advection timescale starting from integrals of the terms
appearing in the radial momentum equation.  This new diagnostic generalizes and
refines the approach by \citet{murphy:13} in which the shock position was
derived starting from an approximated angle-averaged shock jump condition. With
our approach it is possible to quantify, in a rigorous way, the relative
importance of different terms in providing pressure support to the shock. Our
analysis suggests that turbulence plays an indirect role in the revival of the
shock. Rather than directly transporting energy to the shock, turbulence acts as
an effective barrier slowing down the drain of energy from the shock by the
radial advection.

We studied the turbulent energy cascade in the gain region by means of the 3D
power-spectrum of the turbulent velocity $E(k)$.  We find conclusive evidence
that the shallow spectra reported by many investigations are the result of the
numerical bottleneck effect, as previously suggested by \citet{abdikamalov:15}.
In particular, we observe that as resolution increases, the spectra become
progressively steeper. At the highest resolution, the spectrum has a slope
compatible with the $k^{-5/3}$ slope predicted by the classical theory of
Kolmogorov, \eg \cite{pope:00}, and as suggested by local simulations
\citep{radice:15a}.

Unfortunately, resolving the inertial range of neutrino-driven convection
requires resolutions that are not even affordable for a full simulation in our
simplified setup and that could only be employed to simulate a relatively short
time frame ($\simeq 60\ \mathrm{ms}$) starting from a lower resolution
simulation.  This was enough to be able to measure the spectrum of the turbulent
kinetic energy, which we find to have already reached a new equilibrium after
$\sim 3\ \mathrm{ms}$.  However, these $60\ \mathrm{ms}$ of evolution are not
enough to fully assess the ramifications of not resolving the inertial range in
a global simulation. Our simulations appear to be already converged at large
scale, but the difference between resolving and not resolving the inertial range
could become more substantial for models close to the explosion threshold.

At the moment, achieving a resolution sufficient to fully resolve the inertial
range dynamics in global simulations seems to be impossible. At the same time,
the numerical schemes currently adopted for \ac{CCSN} simulations show rather
poor performance for under-resolved turbulent flows \citep{radice:15a}. It is
thus our opinion that performing qualitatively and quantitatively accurate
\ac{CCSN} simulations will require the use of some form of turbulent closure. In
future work, we plan to use the simulation data presented here as a basis to
guide the construction of numerical turbulent closures specialized for \ac{CCSN}
applications.

\section*{Acknowledgements}
We acknowledge helpful discussions with W.~D.~Arnett, A.~Burrows, C.~Meakin,
P.~M\"osta, J.~Murphy, and L.~Roberts. This research was partially supported by
the National Science Foundation under award nos.\ AST-1212170 and PHY-1151197
and by the Sherman Fairchild Foundation. The simulations were performed on the
Caltech compute cluster Zwicky (NSF MRI-R2 award no.\ PHY-0960291), on the NSF
XSEDE network under allocation TG-PHY100033, and on NSF/NCSA BlueWaters under
NSF PRAC award no.\ ACI-1440083.

\bibliography{bibliography/bh_formation_references,%
bibliography/gw_references,%
bibliography/sn_theory_references,%
bibliography/grb_references,%
bibliography/nu_obs_references,%
bibliography/methods_references,%
bibliography/eos_references,%
bibliography/NSNS_NSBH_references,%
bibliography/stellarevolution_references,%
bibliography/nucleosynthesis_references,%
bibliography/gr_references,%
bibliography/nu_interactions_references,%
bibliography/sn_observation_references,%
bibliography/populations_references,%
bibliography/pns_cooling_references,%
bibliography/spectral_photometric_modeling_references,%
bibliography/cs_hpc_references,%
bibliography/numrel_references,%
bibliography/gw_data_analysis_references,%
bibliography/gw_detector_references,%
bibliography/radiation_transport_references,%
bibliography/fluid_dynamics_references,%
bibliography/cosmology_references,%
bibliography/privatebibs/radice}

\begin{appendix}

\section{A. Parametrized Nuclear Dissociation Treatment}
\label{sec:nuclear.dissociation}
Nuclear dissociation is included in a parametrized way using an approach similar
to that of \cite{fernandez:09a, fernandez:09b}, but with some important
differences discussed here.

\cite{fernandez:09a} suggested to parametrize the amount of specific internal
energy lost to nuclear dissociation, $\epsilon_{\mathrm{ND}}$, as a fraction,
$\bar{\epsilon}$, of the free-fall kinetic energy at the initial location of the
shock:
\begin{equation}
  \epsilon_{\mathrm{ND}} = \frac{1}{2} \bar{\epsilon}\,
  \upsilon_{\mathrm{FF}}^2\,,
\end{equation}
where $\upsilon_{\mathrm{FF}}$ is the free-fall velocity at the initial location
of the shock. In the relativistic case this translates to
\begin{equation}
  \epsilon_{\mathrm{ND}} = \bar{\epsilon} ( W_{\mathrm{FF}} - 1 )\,,
\end{equation}
where $W_{\mathrm{FF}}$ is the free-fall Lorentz factor (see Appendix
\ref{sec:standing.accretion.shock}). A typical range of values for
$\bar{\epsilon}$ is $0.2 - 0.4$ \citep{fernandez:09b}.

\cite{fernandez:09a, fernandez:09b} used the nuclear burning module of the FLASH
code \cite{fryxell:00} to simulate nuclear dissociation with the inclusion of an
energy sink term.  This approach is perfectly viable in classical hydrodynamics,
but not in the relativistic case, because, in relativistic hydrodynamics, the
inertia (and momentum) of the fluid depends on the enthalpy and, for this
reason, a sink term in the energy equation would result in an inconsistency with
the shock jump conditions. In our implementation, instead, nuclear dissociation
is included in the equation of state as follows. We model the effect of the
thermal energy lost to nuclear dissociation with a modified gamma-law of the
form
\begin{equation}\label{eq:eos}
  p = (\gamma - 1) \rho \big(\epsilon -
  \epsilon_{\mathrm{ND}}^\ast(\epsilon)\big)\,,
\end{equation}
where
\begin{equation}
  \epsilon^\ast_{\mathrm{ND}}(\epsilon) = \begin{cases}
    0\,, & \textrm{if } \epsilon \leq \epsilon_{\mathrm{ND}}\,, \\
    \eta (\epsilon - \epsilon_{\mathrm{ND}})\,, & \textrm{if }
      \epsilon_{\mathrm{ND}} < \epsilon
      \leq \epsilon_{\mathrm{ND}} \big(\frac{\eta + 1}{\eta}\big)\,, \\
    \epsilon_{\mathrm{ND}}\,, & \textrm{if } \epsilon > \epsilon_{\mathrm{ND}}
      \big(\frac{\eta + 1}{\eta}\big)\,,
  \end{cases}
\end{equation}
and $\eta = 0.95$ is an efficiency parameter needed to ensure that
$p(\rho,\cdot)$ is a one-to-one function (this is needed for the recovery of
$\rho, \upsilon^i$ and $\epsilon$ from the evolved variables at the end of each
iteration during the evolution). Another advantage of this approach, as compared
to the one of \citet{fernandez:09a, fernandez:09b}, is that it does not involve
possibly stiff cooling terms that can give rise to numerical problems.

The results presented in this paper are obtained with $\bar{\epsilon} = 0.3$,
which corresponds to a value of $\epsilon_{\mathrm{ND}} = 0.003$ ($\simeq
2.7\times 10^{18}\ \mathrm{erg}\ \mathrm{g}^{-1}$) for a shock stalled at $100$
Schwarzschild radii of the \ac{PNS} ($\simeq 191\ \km$). We remark that our
results are sensitive to the choice of $\bar{\epsilon}$, because the flow
becomes stable against the development of convection when $\bar{\epsilon}$ is
sufficiently small. The reason is that smaller $\bar{\epsilon}$ result in larger
radial velocities immediately downstream from the shock, which, in turn, prevent
buoyant instabilities from growing into fully-developed convection before being
advected out of the gain region \citep{foglizzo:06}. For more details, we refer
to the studies of \citet{fernandez:09a, fernandez:09b}, and \citet{cardall:15}
that showed the impact of nuclear dissociation on the development of
neutrino-driven convection and \ac{SASI}.

\section{B. Relativistic Standing Accretion Shock Solution}
\label{sec:standing.accretion.shock}
The initial conditions of our simulations represent a stationary standing
accretion shock. Our model is similar to the \ac{NS} accretion models of
\cite{chevalier:89, houck:92} and the \ac{CCSN} model of \cite{janka:01}, which
has been used in many studies of \ac{SASI} and convection, \eg
\citet{blondin:03, foglizzo:06, cardall:15}.

To construct the initial conditions, we solve the equations of relativistic
hydrodynamics \eqref{eq:hydro} on top of a fixed gravitational background
(Equation \ref{eq:metric}) time-independently. Heating, cooling and
nuclear dissociation are also taken into account in the same way as for the
subsequent numerical evolution. The initial conditions are specified by choosing
values for the \ac{PNS} radius, $r_{\mathrm{PNS}}$, the shock radius, $r_s$, the
accretion rate $\dot{M}$, and the heating coefficient $K$. The heating/cooling
normalization coefficient $C$ is then tuned so that the velocity vanishes at the
\ac{PNS} radius.

\subsection{Pre-Shock Flow}
The pre-shock flow is assumed to be cold and free falling, so that the pre-shock
Lorentz factor $W_0$ can be computed from the lapse function at the location
of the shock: $W_0 = \frac{1}{\alpha_0}$. The pre-shock density is computed by
fixing the accretion rate $\dot{M}$:
\begin{equation}\label{eq:preshock.rho}
  \rho_0 = \frac{\dot{M}}{4 \pi r_s^2 W_0 | \upsilon_0^r |}\,,
\end{equation}
where $\upsilon_0^r$ is the radial component of the pre-shock velocity. In
practice, for numerical reasons, to minimize disturbances in the upstream flow,
our initial conditions have a small, but non-zero, pre-shock internal energy
$\epsilon_0$, which we compute from the requirement that the Mach number of the
upstream flow should be equal to $100$.

\subsection{Shock Jump Conditions}
The post-shock density, pressure and velocity can be computed from the
Rankine-Hugoniot conditions of a stationary shock
\begin{subequations}\label{eq:rankine.hugoniot}
\begin{align}
  \rho_1 W_1 \upsilon_1^r &= \rho_0 W_0 \upsilon_0^r\,, \\
  \rho_1 h_1 W_1^2 (\upsilon_1^r)^2 + p_1 \alpha^2 &=
    \rho_0 h_0 W_0^2 (\upsilon_0^r)^2 + p_0 \alpha^2\,, \\
  \rho_1 h_1 W_1^2 \upsilon_1^r &= \rho_0 h_0 W_0^2 \upsilon_0^r\,,
\end{align}
\end{subequations}
where the indices $0$ and $1$ refer to pre- and post-shock quantities,
respectively, and we made use of Equation \eqref{eq:lapse}. Note that the effect
of dissociation is automatically included in these jump conditions, since it is
accounted for in the equation of state.

In the strong shock limit, $\epsilon_0 \ll 1$, and for small post-shock
velocities, $W_1 \simeq 1$, they can be simplified as
\begin{subequations}\label{eq:ss.rankine.hugoniot}
\begin{align}
  \rho_1 \upsilon_1^r &= \rho_0 W_0 \upsilon_0^r\,, \\
  \rho_1 h_1 (\upsilon_1^r)^2 + p_1 \alpha^2 &=
    \rho_0 W_0^2 (\upsilon_0^r)^2\,, \\
  \rho_1 h_1 \upsilon_1^r &= \rho_0 W_0^2 \upsilon_0^r\,.
\end{align}
\end{subequations}
These can easily be solved for the post-shock density,
\begin{equation}
  \rho_1 = \rho_0 W_0 \frac{\upsilon_0^r}{\upsilon_1^r}\,,
\end{equation}
specific internal energy
\begin{equation}\label{eq:postshock.eps}
  \epsilon_1 = \frac{W_0 - 1 + (\gamma - 1)\epsilon_{\mathrm{ND}}}{\gamma}\,,
\end{equation}
and velocity
\begin{equation}
  \upsilon_1^r = \frac{\upsilon_0^r + \sqrt{(\upsilon_0^r)^2 - 4 \psi}}{2}\,,
\end{equation}
where
\begin{equation}
  \psi = \alpha^2 \frac{\gamma-1}{\gamma}
    \frac{W_0-1-\epsilon_{\mathrm{ND}}}{W_0}\,.
\end{equation}
Note that Equation \eqref{eq:postshock.eps} is valid as long as $\bar{\epsilon}$
is sufficiently small so that
\begin{equation}
  \bar{\epsilon}\bigg(\frac{\eta+1}{\eta} - \frac{\gamma-1}{\gamma}\bigg) \leq
  \frac{1}{\gamma}.
\end{equation}
For $\gamma = 4/3$ and $\eta = 0.95$ this means $\bar{\epsilon} \lesssim 0.42$,
which is satisfied since we take $\bar{\epsilon} = 0.3$.

\subsection{Post-Shock Flow}
The initial post-shock flow is computed by looking for a time-independent version
of Equation \eqref{eq:hydro}, with boundary-conditions given by the post-shock
density, velocity and specific internal energy. Note that, as we show later,
the equations are singular at the point where $\upsilon^r=0$ and no boundary
condition is required downstream of the shock. Instead, the normalization
coefficient of the heating/cooling source, $S$, has to be adjusted so that
$\upsilon^r$ vanishes at the surface of the \ac{PNS}.

The first condition that we use is the continuity equation (first part of
Equation \ref{eq:hydro}), that, in the stationary, spherically symmetric case,
is simply
\begin{equation}\label{eq:trivial.continuity}
  r^2 \rho W \upsilon^r = \frac{1}{4\pi}\dot{M}\,.
\end{equation}

The energy equation is rewritten in non-conservation form, by projecting the
second part of Equation \eqref{eq:hydro} along the velocity four-vector, to yield
(see e.g., \citealt{gourgoulhon:06} for a detailed derivation)
\begin{equation}
  u^\mu \nabla_\mu ( \rho \epsilon ) = - (\rho \epsilon + p)\nabla_\mu u^\mu +
  \mathcal{L}\,.
\end{equation}
In the stationary, spherically symmetric case, this becomes
\begin{equation}\label{eq:ode.energy}
  \frac{\dot{M}}{4\pi} \partial_r \epsilon = - 2 p r W \upsilon^r - p r^2
  \partial_r (W \upsilon^r) + \mathcal{L}\,.
\end{equation}

Similarly, the momentum equation is also rewritten in non-conservation form, by
projecting the second part of Equation \eqref{eq:hydro} perpendicularly to
$u^\mu$ to obtain the relativistic Euler equations,
\begin{equation}
  \rho h a_\mu = - \nabla_\mu p - u^\nu \nabla_\nu p u_\mu\,,
\end{equation}
where $a_\mu$ is the relativistic 4-acceleration vector
\begin{equation}
  a_\mu = u^\nu \nabla_\nu u_\mu = u^\nu (\partial_\nu u_\mu -
  \Gamma^{\alpha}_{\phantom{\alpha}\nu\mu} u_\alpha)\,,
\end{equation}
and $\Gamma^{\alpha}_{\phantom{\alpha}\nu\mu}$ are the Christoffel symbols of
the Levi-Civita connection. In our case, the Euler equation reduces to
\begin{equation}\label{eq:ode.momentum}
 \rho h W \upsilon^r A^2 \partial_r (W \upsilon^r) =
  \frac{1}{A^4}\frac{2M}{r^2} \rho h W^2 (\upsilon^r)^2
  - \big[1 + A^2 W^2 (\upsilon^r)^2\big] \partial_r p
  - \frac{A^2}{r^2} M W^2 \big[1 + A^2 (\upsilon^r)^2\big] \rho h\,.
\end{equation}
Note that this equation has a manifest singularity for $\upsilon^r = 0$. In
practice, $\upsilon^r$ is never exactly equal to zero, although tuning $C$
yields very small values of $|\upsilon^r|$ close to the surface of the \ac{PNS},
and we find that the ODE integrator of our choice, the implicit multistep
``MSBDF'' method implemented in the GNU Scientific Library \citep{galassi:09},
is sufficiently robust to handle these equations.

Finally we can substitute the derivative of the pressure in the right-hand-side
of Equation \eqref{eq:ode.momentum} using the \ac{EOS} to find
\begin{equation}\label{eq:ode.momentum2}
\begin{split}
  \Big[\rho h W \upsilon^r A^2  - \frac{\gamma p}{W \upsilon} \big(1 + A^2 W^2
  (\upsilon^r)^2\big) \Big]& \partial_r (W \upsilon^r) = \\
  & \frac{1}{A^4}\frac{2M}{r^2} \rho h W^2 (\upsilon^r)^2
  + \frac{1 + A^2 W^2 (\upsilon^r)^2}{r^2 W \upsilon^r} \big[
    2 \gamma r W \upsilon^r p - (\gamma - 1) \mathcal{L} \big]
  - \frac{A^2}{r^2} M W^2 \big[1 + A^2 (\upsilon^r)^2\big] \rho h\,.
\end{split}
\end{equation}

\begin{figure}
  \includegraphics[width=0.495\columnwidth]{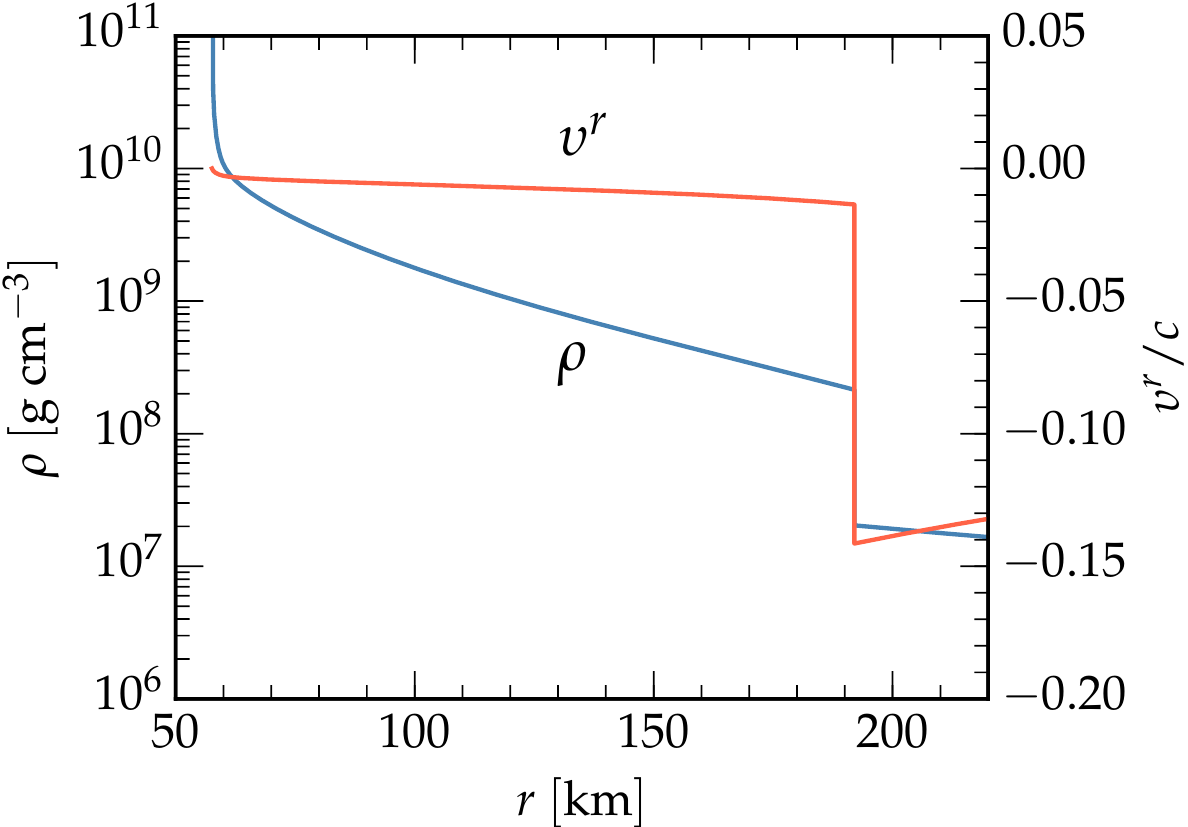}
  \includegraphics[width=0.495\columnwidth]{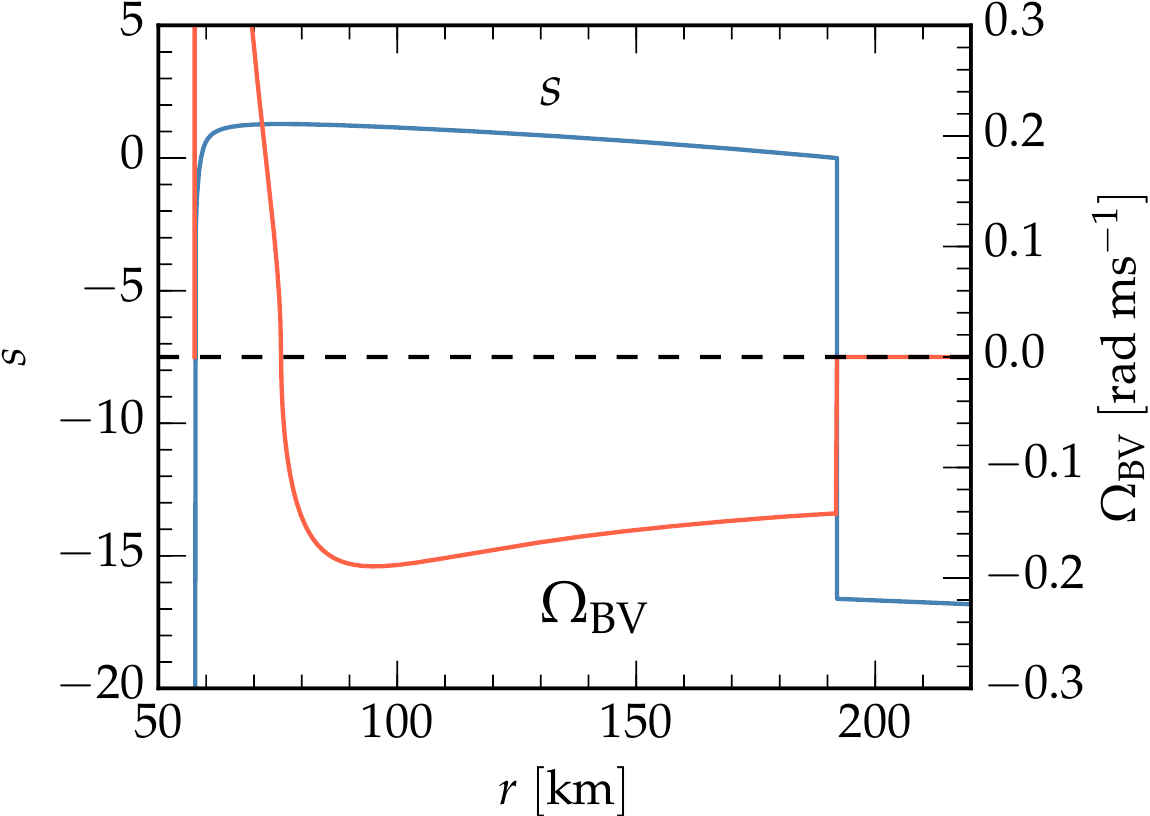}
  \caption{Initial data profiles. \emph{Left panel:} density and velocity.
  \emph{Right panel:} entropy and Brunt-V\"ais\"ala frequency
  $\Omega_{\mathrm{BV}}$ computed as in Equations \eqref{eq:entropy} and
  \eqref{eq:bvfrequency}.} \label{fig:init_data_entropy_omega}
\end{figure}

Equations \eqref{eq:trivial.continuity}, \eqref{eq:ode.energy} and
\eqref{eq:ode.momentum2} are the ones that we solve numerically, together with
the \ac{EOS}, to generate our initial conditions.  Figure
\ref{fig:init_data_entropy_omega}, shows the profile of the entropy and the
Brunt-V\"ais\"ala frequency for the initial conditions used throughout this
study.

\end{appendix}

\acrodef{CCSN}{core-collapse supernovae}
\acrodef{EOS}{equation of state}
\acrodef{NS}{neutron star}
\acrodef{PNS}{proto-neutron star}
\acrodef{SASI}{standing accretion shock instability}
\acrodef{SN}{supernova}

\end{document}